\documentclass[12pt]{article}
\usepackage{amsfonts}
\usepackage{bm}
\usepackage{amsmath}
\usepackage{epsfig}
\usepackage[mathscr]{euscript}

\usepackage{slashed}

\newcommand{\bmat}{\left(\begin{array}}
\newcommand{\emat}{\end{array}\right)}

\def\gtrsim{\mathrel{\raise.3ex\hbox{$>$\kern-.75em\lower1ex\hbox{$\sim$}}}}

\def\a{\alpha}

\def\b{\beta}

\def\-{\hphantom{-}}
\def\ov{\overline}
\def\un{\underline}
\def\s2{\frac{1}{\sqrt2}}

\def\wt{\widetilde}

\def\mg{m_{3/2}}
\def\mg2{m^2_{3/2}}

\def\Dsl{\,\raise.15ex\hbox{/}\mkern-13.5mu D} 

\def\be{\begin{equation}}
\def\ee{\end{equation}}
\def\bea{\begin{eqnarray}}
\def\eea{\end{eqnarray}}

\newcommand{\nn}{\nonumber}

\begin{document}

\pagestyle{plain}

\makeatletter
\@addtoreset{equation}{section}
\makeatother
\renewcommand{\theequation}{\thesection.\arabic{equation}}
\pagestyle{empty}
\begin{center}
\ \

\vskip .5cm
\LARGE{\LARGE\bf Higher-derivative Heterotic Double Field Theory and Classical Double Copy 
 \\[10mm]}
\vskip 0.3cm

\large{Eric Lescano$^\dag$ and Jes\'us A. Rodr\' iguez$^*$
 \\[6mm]}

{\small  $^\dag$ Instituto de Astronom\'ia y F\'isica del Espacio, Universidad de Buenos Aires (UBA-CONICET)\\ [.01 cm]}
{\small\it Ciudad Universitaria, Pabell\'on IAFE, 1428 Buenos Aires, Argentina\\ [.3 cm]}
{\small  $^*$ Departamento de F\'isica, FCEyN, Universidad de Buenos Aires (UBA-CONICET) \\ [.01 cm]}
{\small\it Ciudad Universitaria, Pabell\'on 1, 1428 Buenos Aires, Argentina\\ [.5 cm]}

{\small \verb"elescano@iafe.uba.ar, jarodriguez@df.uba.ar"}\\[1cm]

\small{\bf Abstract} \\[0.5cm]\end{center}
 
The generalized Kerr-Schild ansatz (GKSA) is a powerful tool for constructing exact solutions in Double Field Theory (DFT). In this paper we focus in the heterotic formulation of DFT, considering up to four-derivative terms in the action principle, while the field content is perturbed by the GKSA. We study the inclusion of the generalized version of the Green-Schwarz mechanism to this setup, in order to reproduce the low energy effective heterotic supergravity upon parametrization. This formalism reproduces higher-derivative heterotic background solutions where the metric tensor and Kalb-Ramond field are perturbed by a pair of null vectors. Next we study higher-derivative contributions to the classical double copy structure. After a suitable identification of the null vectors with a pair of $U(1)$ gauge fields, the dynamics is given by a pair of Maxwell equations plus higher derivative corrections in agreement with the KLT relation.

\newpage
\setcounter{page}{1}
\pagestyle{plain}
\renewcommand{\thefootnote}{\arabic{footnote}}
\setcounter{footnote}{0}

\tableofcontents
\newpage

\section{Introduction}
Double Field Theory (DFT) \cite{Siegel,DFT} is a duality symmetric formalism that can be understood as a generalization of $D$-dimensional Riemannian geometry, manifestly invariant under the action of $O(D,D)$. This group is closely related with an exact symmetry of String Theory \cite{Ashoke}. One of the most exciting features of this formalism is that the space-time must be doubled to accomplish $O(D,D)$ as a global symmetry of the theory \cite{ReviewDFT}. The generalized coordinates of the double space $X^{M}=(x^{\mu},\tilde{x}_{\mu})$ are in the fundamental representation of $O(D,D)$, where $\tilde{x}_{\mu}$ is the extra set of coordinates and $M=0,\dots,2D-1$. However, the theory is constrained by the section condition (or strong constraint), which effectively removes the dependence on $\tilde{x}_{\mu}$.

The fundamental bosonic fields consist of a symmetric generalized tensor ${H}_{M N}(X)$ and a generalized scalar $d_{o}(X)$. While the former is a multiplet of the duality group, the latter is an invariant, and they are usually referred as the generalized metric and the generalized dilaton. On the other hand, DFT can easily describe the low energy limit of heterotic string theory \cite{HetDFT}. This formalism requires a generalized frame \cite{FrameDFT} instead of a generalized metric to embed the gravitational degrees of freedom of the theory. A similar approach is the flux formulation of DFT \cite{flux}. 

One important issue that can be studied with DFT is the incorporation of higher derivative terms in supergravity frameworks. In \cite{Tduality} a biparametric family of higher-derivative duality invariant theories was presented. For particular values of the parameters, the higher-derivative contributions reproduce the $\alpha'$-corrections of the heterotic supergravity, also studied in \cite{alpha}.

In this work we are interested in considering $H_{M N}$ and $d_{o}$ as arbitrary background fields that can be perturbed. We focus in the generalized Kerr-Schild ansatz (GKSA) introduced in \cite{KL}, which is an exact and linear perturbation of the generalized background metric given by 
\bea
{\cal H}_{MN} = H_{MN} + \kappa \bar{K}_{M} K_{N} + \kappa {K}_{M} \bar{K}_{N}  \, ,
\eea
where $\kappa$ is an arbitrary parameter that allows to quantify the order of the perturbations and $\bar{K}_{M}= \bar{P}_{M}{}^{N} \bar{K}_{N}$ and $K_{M}= {P}_{M}{}^{N} {K}_{N}$ are a pair of generalized null vectors 
\bea
\eta^{MN} \bar{K}_{M} \bar{K}_{N} & = & \eta^{MN} K_{M} K_{N} =  \eta^{MN} \bar{K}_{M} K_{N} = 0 \, \, ,
\eea 
where $\eta_{M N}$ is an $O(D,D)$ invariant metric and $\bar{P}_{MN}=\frac12(\eta_{MN} + {H}_{MN})$ and ${P}_{MN}=\frac12(\eta_{MN} - {H}_{MN})$ are used to project the $O(D,D)$ indices. This ansatz (plus a generalized dilaton perturbation) is the duality invariant analogous of the ordinary Kerr-Schild ansatz \cite{KS} \cite{KSapp}, and the inclusion of a pair of generalized null vectors is closely related with the chiral structure of DFT. This duality invariant ansatz was recently used in different context as exceptional field theory \cite{EFTKS}, supersymmetry \cite{AleyEric}, among others \cite{GKSAo}.   

In \cite{HetKL} a consistent way to impose the GKSA in a two-derivative heterotic DFT framework was described. In that work the authors considered a generalized metric approach to recover the leading order contributions. Here we extend the formalism by considering higher-derivative terms in the flux formulation of DFT. We use a systematic method for obtaining these corrections, closely related with \cite{gbdr}. Higher-derivative contributions requiere agreement between the ansatz and the generalized version of the Green-Schwarz mechanism as we show. 

As an application we study higher-derivative contributions to the classical double copy \cite{KLT} \cite{BCJ} \cite{Dcopy} \footnote{See also \cite{Dcopyothers} for recent approaches to this topic.} in a heterotic supergravity background. We consider,
\bea
g^{\mu \nu} & = & g_{o}^{\mu \nu} + \kappa l^{(\mu} \bar{l}^{\nu)} \, , \\
g_{\mu \nu} & = & g_{o\mu \nu} - \wt{\kappa}  \bar l_{(\mu} l_{\nu)} \, ,
\eea
where $\wt{\kappa}=\frac{2\kappa}{2+\kappa(l\cdot\bar l)}$ and $l$ and $\bar{l}$ are a pair of null vectors with respect to the background metric. We compute the $\rm{Riem}^2$ contributions and we obtain the leading four-derivative corrections to the single copy, given by $\kappa$ corrections to the Maxwell-like equations that described the gravity sector after identifying the null vectors with a pair of $U(1)$ gauge fields, in agreement with the KLT relation for heterotic string theory.

This work is organized as follows: In Section \ref{Sugra} we introduce the field content, the symmetries and the action principle of the low energy effective heterotic supergravity, considering up to four-derivative terms in the action principle. Section \ref{GKS} is dedicated to explore the extension of the GKSA to the flux formalism of DFT. First we review the generalized metric formalism and then we discuss the generalized frame formalism that is necessary to construct the generalized fluxes. In Section \ref{HDFT} we start by considering multiplets of $O(D,D+K)$, and we perform a suitable breaking to $O(D,D)$ by identifying the extra gauge field with a particular flux component. With this method we obtain the higher-derivative extension to DFT. Here we choose the free parameters of the construction to match with the heterotic DFT formulation. In Section \ref{Het} we parametrize the theory in terms of the field content of  heterotic supergravity. We discuss about the tension between the perturbation of the gravitational sector in terms of a pair of null vector with respect to the absence of a perturbation for the gauge sector. Then we apply our formalism in Section \ref{CDC} to explore higher-derivative corrections to the heterotic Classical Double Copy. Finally, in Section \ref{Con}, we present the conclusions of the work.

\section{Higher-derivative heterotic supergravity}
\label{Sugra}

In the first part of this section we review the $D=10$ heterotic supergravity considering $\a$ and $\beta$ contributions, according to \cite{BdR} \footnote{We take $\a=1$ and $\b=1$ to simplify notation. Conventions for the field content follows \cite{AleyEric}.}. Then we impose the supergravity version of the GKSA to perturbe the background fields.  

\subsection{Action and field content for background fields}

The low energy effective action that describes $D=10$ heterotic string theory to first order in $\alpha'$ is
\bea
S=\int_{M_{10}} e^{-2\phi_{o}} \Big(R_{o} - 4 \partial_{\mu}\phi_o \partial^{\mu}\phi_o - \frac{1}{12} \hat{H}_{o \mu \nu \rho} \hat H_{o}^{\mu \nu \rho} - \frac14 (F_{o\mu \nu i} F_{o}^{\mu \nu i} + R_{o \mu \nu}^{(-)}{}^{a b} R_{o}^{(-) \mu \nu}{}_{a b}) \Big) \, , 
\label{alphasugra}
\eea
where $\mu=0,\dots,9$, $a=0,\dots,9$, $i=1,\dots,n$ with $n$ the dimension of the Yang-Mills group, typically $n=496$. The $10$-dimensional field content consists of a background metric $g_{o \mu \nu}=e_{o \mu a} \eta^{a b} e_{o \nu b}$, the background Kalb-Ramond field $b_{o \mu \nu}$, the background gauge field $A_{o \mu i}$ and the background dilaton $\phi_{o}$. The action (\ref{alphasugra}) is written in terms of the curvatures of the previous fields, \textit{i.e.},
\bea
\hat H_{o \mu\nu\rho}&=&3\left[\partial_{[\mu}b_{o\nu\rho]} - \left(A_{o[\mu}^i\partial_\nu A_{o\rho]i} - \frac{1}{3}f_{ijk}A_{o\mu}^iA_{o\nu}^jA_{o\rho}^k\right) \right. \nn \\ && \left. - \left(\partial_{[\mu} \Omega^{(-)cd}_{o\nu} \Omega^{(-)}_{o \rho]cd}+\frac 23 \Omega^{(-)ab}_{o\mu} \Omega^{(-)}_{o\nu bc} \Omega^{(-)}_{o\rho}{}^{c}{}_{a}\right)\right] \, , \nn \\
F_{o\mu\nu}^i&=&2\partial_{[\mu}A^i_{o\nu]}-f^i{}_{jk}A_{o\mu}^j A_{o\nu}^k \, ,\nn \\
R^{(-)}_{o\mu\nu ab}&=& \left(-2\partial_{[\mu} w_{o\nu]ab} + 2w_{o[\mu |a}{}^{{c}} w_{o|\nu] cb} \right)
\, ,
\eea
where the spin and Hull connections are defined as
\bea
w_{o\mu a}{}^{b}& = & -e_{o}^\nu{}_a \partial_\mu e_{o\nu}{}^b + \Gamma_o^\sigma{}_{\mu \nu} e_{o\sigma}{}^b e_{o}^\nu{}_a \, , \\
{\Omega}^{(-)}_{o\mu ab} & = & w_{o\mu ab} - \frac12 \hat{H}_{o\mu \nu \rho} e_{o}^{\nu}{}_{a} e_{o}^{\rho}{}_{b} \, , 
\eea
and the Christoffel connection is
\bea
\Gamma_{o}^{\sigma}{}_{\mu\nu} = \frac{1}{2}g_{o}^{\sigma\rho}\left(\partial_{\mu}g_{o\nu\rho} + \partial_{\nu}g_{o\mu\rho} - \partial_{\rho}g_{o\mu\nu}\right)\, .
\eea
 
The action principle (\ref{alphasugra}) is invariant under Lorentz transformations. These transformations acting on a generic vector $V_{a}$ reads
\bea
\delta_{\Lambda} V_{a} = V_{b} \Lambda^{b}{}_{a} \, ,
\label{stransf}
\eea
where $\Lambda_{a b}=-\Lambda_{b a}$ is the Lorentz parameter. Contraction of indices in (\ref{stransf}) make use of the Lorentz metric $\eta_{a b}$. The invariance of (\ref{alphasugra}) requires that the Kalb-Ramond field transforms with a non-covariant transformation, namely,
\bea
\delta_{\Lambda} b_{o \mu \nu} = - \Omega^{(-)}_{o[\mu}{}^{a b} \partial_{\nu]} \Lambda_{a b}.
\label{bgs}
\eea
The previous transformation is known as the Green-Schwarz mechanism and it is a higher-derivative transformation that cannot be removed with field redefinitions.  

The equations of motion for this setup are given by
\bea
\Delta \phi_{o} & =  & R_{o} + 4 g_{o}^{\mu\nu} (\nabla_\mu  \nabla_\nu \phi_{o}  - \partial_\mu \phi_{o} \partial_\nu \phi_{o}) - \frac 1 {12} g_{o}^{\mu\sigma} g_{o}^{\nu\tau} g_{o}^{\rho\xi} \hat H_{o\mu\nu\rho} \hat H_{o\sigma\tau\xi} \label{DeltaPhi} \\ && - \frac {1} 4g_{o}^{\mu\rho} g_{o}^{\nu\sigma} F_{o\mu\nu}{}^i F_{o\rho \sigma i}  - \frac {1} 4g_{o}^{\mu\rho} g_{o}^{\nu\sigma}  R^{(-)}_{o\mu\nu a b}  R^{(-)}_{o\rho \sigma}{}^{a b} \ = \ 0\, ,\nn\\
\Delta g_{o\mu\nu} &=& R_{o\mu\nu} + 2 \nabla_\mu \nabla_\nu \phi_{o} - \frac 1 4g_{o}^{\sigma \tau} g_{o}^{\lambda \xi} \hat H_{o\sigma\lambda\mu} \hat H_{o\tau\xi\nu}\nn\\
&& - \frac {1} 2 g_{o}^{\sigma \tau } F_{o\sigma \mu i} F_{o\tau \nu}{}^i - \frac {1} 2 g_{o}^{\sigma \tau } R^{(-)}_{o\sigma \mu a b}  R^{(-)}_{o\tau \nu}{}^{a b} \ = \ 0\, , \label{Deltag}\\
\Delta b_{o \mu\nu} &=& g_{o}^{\rho \sigma} \nabla_\rho \left(e^{-2\phi_{o}}\hat H_{o\mu\nu\sigma}\right) \ = \ 0\, ,\label{DeltaB}\\
\Delta A_{o\nu}{}^i &=& g_{o}^{\rho \mu } \nabla^{(+,A)}_\rho  \left(e^{-2\phi_{o}} F_{o\mu\nu}{}^i\right) \ = \ 0\, , \label{eqA}
\eea
where we have defined
\bea
\nabla^{(+,A)}_\rho F_{o\mu\nu}{}^i &=& \partial_\rho F_{o\mu\nu}{}^i - \Gamma^{(+)}_{o\rho\mu}{}^\sigma F_{o\sigma\nu}{}^i - \Gamma^{(+)}_{o\rho\nu}{}^\sigma F_{o\mu\sigma}{}^i - f_{jk}{}^i A_{o\rho}{}^j F_{o\mu\nu}{}^k\, ,\\
\Gamma^{(+) \rho}_{o\mu\nu} &=& \Gamma^\rho_{o\mu\nu} + \frac 1 2 \hat H_{o\mu\nu\sigma} g_{o}^{\sigma \rho}\, ,\eea
which covariantizes the derivative with respect to ten dimensional diffeomorphisms and gauge transformations using the Christoffel and gauge connection respectively. Here we mention that the equations (\ref{DeltaPhi})-(\ref{eqA}) are not strictly the ones obtained from variations of the action with respect to the fundamental fields, but combinations of them.

\subsection{Perturbing the background}

Now we are interested in imposing the supergravity version of the GKSA on the previous formulation. We do not consider perturbations of the gauge field, \textit{i.e.}, $A_{\mu i}=A_{o\mu i}$. The inverse of the 10-dimensional background metric is perturbed as
\bea
g^{\mu \nu} & = & g_{o}^{\mu \nu} + \kappa l^{(\mu} \bar{l}^{\nu)} \, ,
\label{metricp}
\eea
where $l_{\mu}$ and $\bar{l}_{\mu}$ are null vectors with respect to $g^{\mu \nu}$ and $g_{o}^{\mu \nu}$, \textit{i.e.},
\bea
l_{\mu} l_{\nu} g^{\mu \nu} = l_{\mu} g_{o}^{\mu \nu} l_{\nu}  = 0 \, , \\
\bar l_{\mu} \bar l_{\nu} g^{\mu \nu} = \bar l_{\mu}  g_{o}{}^{\mu \nu} \bar l_{\nu} = 0 \, .
\eea

The previous objects also satisfy the following relations
\bea
\bar{l}^{\nu} \nabla_{o\nu}l_\mu =  0 \, , \quad {l}^{\nu} \nabla_{o\nu}{\bar l}_\mu =  0 \, , 
\label{geode}
\eea
which reduce to the standard geodesic conditions when $l$ and $\bar{l}$ are identified \cite{KL}. The perturbation of the Kalb-Ramond field is given by
\bea
 b_{\mu \nu} & = & b_{o \mu \nu} - \wt{\kappa} l_{[\mu} \bar{l}_{\nu]} \, ,
\label{bp}
\eea
where $\wt{\kappa}=\frac{2\kappa}{2+\kappa(l\cdot\bar l)}$ and the dilaton is perturbed as showed in \cite{AleyEric}. Moreover, the first order Lorentz transformation for the exact $b$-field now takes the following form
\bea
\delta_{\Lambda} b_{\mu \nu} & = & - \Omega^{(-)}_{[\mu}{}^{a b} \partial_{\nu]} \Lambda_{a b} \, ,
\label{transf1}
\eea
while the other exact fields are Lorentz invariant.

\section{Double Field Theory and the generalized Kerr-Schild Ansatz}
\label{GKS}

\subsection{Generalized metric formulation}

In this section we review DFT and the GKSA following the conventions of \cite{AleyEric}.
The GKSA was formulated in \cite{KL} as an exact and linear perturbation of the generalized background metric $H_{M N}$ ($M,N=0, \dots, 2D-1$) and an exact perturbation of the generalized background dilaton $d_{o}$. We work with arbitrary $D$ until parametrization. 

Since the generalized metric is an $O(D,D)$ element, its perturbation has the following form
\bea
{\cal H}_{MN} = H_{MN} + \kappa (\bar{K}_{M} K_{N} + {K}_{M} \bar{K}_{N} ) \, ,
\label{DFTKS}
\eea
where $\bar{K}_{M}= \bar{P}_{M}{}^{N} \bar{K}_{N}$ and $K_{M}= {P}_{M}{}^{N} {K}_{N}$ are a pair of generalized null vectors 
\bea
\eta^{MN} \bar{K}_{M} \bar{K}_{N} & = & \eta^{MN} K_{M} K_{N} =  \eta^{MN} \bar{K}_{M} K_{N} = 0 \, .
\label{nulldft}
\eea 
According to (\ref{DFTKS}), the DFT projectors are
\bea
{\cal P}_{MN} &=& P_{MN} - \frac12 \kappa (\bar{K}_{M} K_{N} + {K}_{M} \bar{K}_{N}) \nn \\ 
\bar{\cal P}_{MN} &=& \bar{P}_{MN} + \frac12 \kappa (\bar{K}_{M} K_{N} + {K}_{M} \bar{K}_{N}) \, .
\eea
Each $O(D,D)$ multiplet can be written as a sum over its projections,
\bea
V_{M} = P_{M}{}^{N} V_{N} + {\bar P}_{M}{}^{N} V_{N} = V_{\underline M} + V_{\overline M} \, ,
\eea
where $V_{M}$ is a generic double vector. When we use the underline and overline notation, we consider the background projectors $P_{MN}$ and ${\bar P}_{MN}$.

The generalized background dilaton can be perturbed with a generic $\kappa$ expansion,
\be
d = d_{o} + \kappa f\, , \qquad f = \sum_{n=0}^{\infty}\kappa^{n}f_{n} \, ,
\ee
with $n\geq 0$.

Mimicking the ordinary Kerr-Schild ansatz, the generalized vectors $K_{M}$, $\bar{K}_{M}$ and $f$ obey some conditions in order to produce finite deformations in the DFT action and EOM's. If we consider a generic double vector $V_{N}$, the covariant derivative can be defined as
\bea
\nabla_{M} V_{N} = \partial_{M} V_{N} - \Gamma_{MN}{}^{P} V_{P} \, , 
\eea
where $\Gamma_{MNP}$ is the generalized affine connection. 
Demanding 
\bea
\nabla_{M}{\cal H}_{NP}&=&0 \, , \quad \nabla_{M}{\cal \eta}_{NP}=0 \, ,  
\eea
and a vanishing generalized torsion 
\bea
\Gamma_{[MNP]}=0 \, ,
\label{torsion}
\eea
the following projections of $\Gamma_{MNP}=-\Gamma_{MPN}$ are well-defined and can be perturbed,
\bea
\Gamma_{\underline M \underline N \overline Q} & = &  -\bar{\cal P}_{Q}{}^{R} {\cal P}_{M}{}^{S} \partial_{S}{\cal P}_{R N}  \, , \quad \Gamma_{\overline M \overline N \underline Q} =   \bar{\cal P}_{N}{}^{R} {\bar {\cal P}}_{M}{}^{S}\partial_{S} {\cal P}_{R Q}  \, ,  \nn \\
\Gamma_{\underline M \overline N \overline Q} & = &  2 \bar{\cal P}_{[N}{}^{R}  {\bar {\cal P}}_{Q]}{}^{S} \partial_{S} {\cal P}_{R M}  \, , \quad \Gamma_{\overline M \underline N \underline Q} =  2 \bar{\cal P}_{M}{}^{R} {\cal P}_{[N}{}^{S}\partial_{S} {\cal P}_{Q]R} \, .
\eea

Similarly to Riemannian geometry, the generalized Ricci scalar and the generalized Ricci tensor can be constructed from different (determined) projections of the generalized affine connection. Following the original construction of the GKSA we impose,
\bea
{\bar K}^{P} \partial_{P} K^{M} + K_{P} \partial^{M}{\bar K}^{P} - K^{P} \partial_{P}{\bar K}^{M} & = & 0 \, , \nn \\ {K}^{P} \partial_{P} {\bar K}^{M} + {\bar K}_{P} \partial^{M}{K}^{P} - {\bar K}^{P} \partial_{P}{K}^{M} & = & 0 \, , 
\label{geodesic1}
\eea
and
\bea
K^{M} \partial_{M}f = {\bar K}^{M} \partial_{M}f = 0 \, .
\label{geodesic2}
\eea
Using (\ref{torsion}), we can change $\partial \rightarrow \nabla$ in (\ref{geodesic1}) obtaining, 
\bea
{\bar K}^{P} \nabla_{P} K^{M} + K_{P} \nabla^{M}{\bar K}^{P} - K^{P} \nabla_{P}{\bar K}^{M} & = & 0 \, , \nn \\ {K}^{P} \nabla_{ P} {\bar K}^{M} + {\bar K}_{P} \nabla^{M}{K}^{P} - {\bar K}^{P} \nabla_{P}{K}^{M} & = & 0 \, . 
\label{geodesic11}
\eea
In the next part we explore how the previous conditions appear in the flux formalism of DFT \cite{flux}. Then we explicitly compute the EOM's of the field content of DFT when we impose the GKSA in this formalism.

\subsection{Generalized flux formulation}
The generalized flux formulation of DFT is closely related with the generalized frame formulation introduced in \cite{FrameDFT}. The latter is compatible with the GKSA if we consider perturbations of the form,
\bea
{\cal E}_{M}{}^{\ov A} = E_{M}{}^{\ov A} + \frac12 \kappa E_{M}{}^{\underline B} K_{\underline B} {\bar K}^{\overline A} \, , \nn \\ {\cal E}_{M}{}^{\underline A} = E_{M}{}^{\underline A} - \frac12 \kappa E_{M}{}^{\ov B} {\bar K}_{\overline B} K^{\underline A}\, ,
\label{GKSA}
\eea
where $K_{A} = {\cal E}^{M}{}_{\underline A} K_{M}={E}^{M}{}_{\underline A} K_{M}$ and $\bar{K}_{A} = {\cal E}^{M}{}_{\overline A} \bar{K}_{M}=E^{M}{}_{\overline A} \bar{K}_{M}$ and ${\cal E}_{MA}$ is an $O(D,D)/O(D-1,1)_{L} \times O(1,D-1)_{R}$ frame. In this formulation $\underline A= 0, \dots, D-1$ and $\overline A=0, \dots, D-1$ are $O(D-1,1)_{L}$ and $O(1,D-1)_{R}$ indices, respectively. 

We can define flat invariant projectors as follows,
\bea
{\cal P}_{A B} & = & {\cal E}_{M \underline A} {\cal E}^{M}{}_{\underline B} = P_{A B} \, , \nn \\ \bar{\cal P}_{{A B}} & = & {\cal E}_{M \ov A} {\cal E}^{M}{}_{\ov B} = \bar P_{{A B}} \, ,
\eea
where $P_{AB}=E_{M \underline A} E^{M}{}_{\underline B}$ and $\bar{P}_{AB}=E_{M \overline A} E^{M}{}_{\overline B}$ are the standard DFT flat projectors. Using these projectors we can construct two invariant metrics,
\bea
\eta_{AB} = {\cal E}_{M A}\eta^{MN} {\cal E}_{N B} = E_{M A}\eta^{MN} E_{N B} \, , \\
{H}_{AB} = {\cal E}_{MA} {\cal H}^{MN} {\cal E}_{N B} = E_{MA} {H}^{MN}E_{N B} \, .
\eea

The flat covariant derivative acting on a generic vector $V_{B}$ is
\bea
{\cal D}_{A} V_{B} = {\cal E}_{A} V_{B} + {\cal W}_{AB}{}^{C} V_{C} \, , 
\eea
where ${\cal E}_{A} = \sqrt{2} {\cal E}^{M}{}_{A} \partial_{M}$ and ${\cal W}_{AB}{}^{C}$ is the generalized spin connection that satisfies
\be
{\cal W}_{ABC} = - {\cal W}_{ACB}\, \qquad \mathrm{and} \qquad {\cal W}_{A\overline B \underline C} = {\cal W}_{A\underline B \overline C} = 0 \, .
\ee
With the help of the generalized frames we can construct the generalized fluxes, which are defined as
\bea
{\cal F}_{ABC} & = & 3 {\cal E}_{[A}({\cal E}^{M}{}_{B}){\cal E}_{M C]} \, , \nn \\ {\cal F}_{A} & = & \sqrt{2}e^{2d}\partial_{M}\left({\cal E}^{M}{}_{A}e^{-2d}\right)  \,.
\eea

In the flux formulation of DFT, conditions (\ref{geodesic1}) and (\ref{geodesic2}) become
\bea
K^{\underline A} E_{\underline A} {\bar K}^{\overline C} + {K}^{\un A} \bar K^{\ov B} F_{\un A \ov B}{}^{\ov C} & = & 0 \, , \nn \\
{\bar K}^{\overline A} E_{\overline A} {K}^{\underline C} + {\bar K}^{\ov A}  K^{\un B} F_{\ov A \un B}{}^{\un C} & = & 0 \, ,
\label{flatgeo1}
\eea
and
\bea
K^{\underline A} E_{\underline A} f = {\bar K}^{A} E_{\overline A} f = 0 \, . 
\label{flatgeo2}
\eea
It is straightforward to check that the previous conditions are double Lorentz invariant using
\bea
\delta_{\Gamma}{\cal E}_{M A} = {\cal E}_{M}{}^{B} \Gamma_{B A} \, , \quad \delta_{\Gamma}{E}_{M A} = {E}_{M}{}^{B} \Gamma_{B A}  
\label{Lorentz}
\eea 
where $\Gamma_{A B} = - \Gamma_{B A}$ is the double Lorentz parameter. 

Only the totally antisymmetric  and trace parts of ${\cal W}_{ABC}$ can be determined in terms of ${\cal E}_{M}{}^{A}$  and $d$,
\bea
{\cal W}_{[ABC]} & = & -\frac13{\cal F}_{ABC}\, , \\
{\cal W}_{BA}{}^{B} & = &  - {\cal F}_{A}\, ,
\eea
the latter arising from partial integration with the dilaton density. Using these identifications, conditions (\ref{flatgeo1}) and (\ref{flatgeo2}) can be written as 
\bea
K_{\underline A} D^{\underline A} {\bar K}^{\overline B}  & = &  \bar K_{\overline A} { D}^{\overline A} {K}^{\underline B} =  0 \, , \nn \\ K_{\underline A} {D}^{\underline A} f & = & K_{\overline A} {D}^{\overline A}f = 0 \, ,
\eea
where $D_{A}$ is the background covariant derivative. As we mentioned before, the generalized Ricci scalar and the generalized Ricci tensor are completely determined in terms of the degrees of freedom of DFT and, particularly, can be written in terms of different projections of the fluxes,
\bea
\label{GRicci_scalar}
{\cal R} & = & 2{\cal E}_{\underline{A}}{\cal F}^{\underline{A}} + {\cal F}_{\underline{A}}{\cal F}^{\underline{A}} - \frac16 {\cal F}_{\underline{ABC}} {\cal F}^{\underline{ABC}} - \frac12{\cal F}_{\ov{A}\underline{BC}}{\cal F}^{\ov{A}\underline{BC}} \, , \\
{\cal R}_{\ov{A}\un{B}} & = & {\cal E}_{\ov{A}}{\cal F}_{\un{B}} - {\cal E}_{\un{C}}{\cal F}_{\ov{A}\un{B}}{}^{\un{C}} + {\cal F}_{\un{C}\ov{DA}}{\cal F}^{\ov{D}}{}_{\un{B}}{}^{\un{C}} - {\cal F}_{\un{C}}{\cal F}_{\ov{A}\un{B}}{}^{\un{C}} \, .
\label{GRicci_tensor}
\eea

The previous projections of the fluxes can be computed using (\ref{GKSA}) and imposing (\ref{flatgeo1}) and (\ref{flatgeo2}), 
\bea
\label{constrained_fluxes0}
{\cal F}_{\un{ABC}} & = & F_{\un{ABC}} -\frac{3}{2}\kappa\ov{K}^{\ov{D}}K_{[\un{A}}F_{\un{BC}]\ov{D}}\, , \\
{\cal F}_{\un{A}\ov{BC}} & = & F_{\un{A}\ov{BC}} + \kappa\left(\ov{K}{}_{[\ov{C}}D{}_{\ov{B}]}K_{\un{A}} + K_{\un{A}}E_{[\ov{B}}\ov{K}_{\ov{C}]} - \frac{1}{2}\ov{K}^{\ov{D}}K_{\un{A}}F_{\ov{DBC}}\right)\, , \\
{\cal F}_{\ov{A}\un{BC}} & = & F_{\ov{A}\un{BC}} - \kappa\left(K_{[\un{C}}D_{\un{B}]}\ov{K}_{\ov{A}} + \ov{K}_{\ov{A}}E_{[\un{B}}K_{\un{C}]} - \frac{1}{2}K^{\un{D}}\ov{K}_{\ov{A}}F_{\un{DBC}}\right)\, , \\
{\cal F}_{\un{A}} & = & F_{\un{A}} - \frac{1}{2}\kappa\left(K_{\un{A}}D_{\ov{B}}\ov{K}^{\ov{B}} + F_{\ov{B}\un{AC}}\ov{K}^{\ov{B}}K^{\un{C}} + 4D_{\un{A}}f\right)\, .
\label{constrained_fluxes}
\eea
Replacing the previous expressions in (\ref{GRicci_scalar}) the generalized Ricci scalar can be written as
\bea
{\cal R} & = & R + \kappa\left[- K_{\un{A}}\ov{K}^{\ov{B}}E_{\ov{B}}F^{\un{A}} - D_{\un{A}}\left(K^{\un{A}}D_{\ov{B}}\ov{K}^{\ov{B}} + F_{\ov{B}}{}^{\un{AC}}\ov{K}^{\ov{B}}K_{\un{C}}\right) + F^{\ov{A}\un{BC}}K_{\un{C}}D_{\un{B}}\ov{K}_{\ov{A}}\right.\, \nn \\
& & \ \ \ \ \ \ \ \ \ \ \left. + F^{\ov{A}\un{BC}}\ov{K}_{\ov{A}}E_{\un{B}}K_{\un{C}} - 4D_{\un{A}}D^{\un{A}}f\right] + \kappa^{2}\left[4E_{\un{A}}f E^{\un{A}}f\right]\,,
\eea
and therefore in the case $f=\mathrm{const.}$, the generalized Ricci scalar can be linearized. 

With a similar procedure the generalized Ricci tensor can be written as,
\bea
{\cal R}_{\ov{A}\un{B}} = {R}_{\ov{A}\un{B}} + \kappa {\cal R}_{(\kappa)\ov{A}\un{B}} + \kappa^2 {\cal R}_{(\kappa^{2})\ov{A}\un{B}} \, , 
\eea
where 
\bea
{\cal R}_{(\kappa)\ov{A}\un{B}} & = & - \frac{1}{2}D_{\ov{A}}\left(K_{\un{B}}D_{\ov{C}}\ov{K}^{\ov{C}}\right) + \frac{1}{2}E^{\un{C}}\left(K_{\un{C}}D_{\un{B}}\ov{K}_{\ov{A}}\right) - \frac{1}{2}E^{\un{C}}\left(K_{\un{B}}D_{\un{C}}\ov{K}_{\ov{A}}\right) + \frac{1}{2}E^{\un{C}}\left(\ov{K}_{\ov{A}}E_{\un{B}}K_{\un{C}}\right)\, \nn \\
& & - \frac{1}{2}E^{\un{C}}\left(\ov{K}_{\ov{A}}E_{\un{C}}K_{\un{B}}\right) - \frac{1}{2}\ov{K}^{\ov{D}}K^{\un{C}}E_{\ov{A}}F_{\ov{D}\un{BC}} - \frac{1}{2}E_{\ov{A}}\ov{K}^{\ov{D}}K^{\un{C}}F_{\ov{D}\un{BC}} - \frac{1}{2}\ov{K}^{\ov{D}}E_{\ov{A}}K^{\un{C}}F_{\ov{D}\un{BC}}\, \nn \\
& & - \frac{1}{2}E^{\un{C}}K^{\un{D}}\ov{K}_{\ov{A}}F_{\un{DBC}} - \frac{1}{2}K^{\un{D}}E^{\un{C}}\ov{K}_{\ov{A}}F_{\un{DBC}} - \frac{1}{2}K^{\un{D}}\ov{K}_{\ov{A}}E^{\un{C}}F_{\un{DBC}} + \frac{1}{2}\ov{K}^{\ov{D}}K^{\un{C}}E_{\ov{D}}F_{\ov{A}\un{BC}}\, \nn \\
& & - \frac{1}{2}K_{\un{C}}D_{\un{B}}\ov{K}_{\ov{E}}F^{\un{C}\ov{E}}{}_{\ov{A}} + \frac{1}{2}K_{\un{B}}D_{\un{C}}\ov{K}_{\ov{E}}F^{\un{C}\ov{E}}{}_{\ov{A}} - \frac{1}{2}\ov{K}_{\ov{E}}E_{\un{B}}K_{\un{C}}F^{\un{C}\ov{E}}{}_{\ov{A}} + \frac{1}{2}\ov{K}_{\ov{E}}E_{\un{C}}K_{\un{B}}F^{\un{C}\ov{E}}{}_{\ov{A}}\, \nn \\
& & + \frac{1}{2}\ov{K}{}_{\ov{A}}D{}_{\ov{D}}K_{\un{C}}F_{\un{B}}{}^{\un{C}\ov{D}} - \frac{1}{2}\ov{K}{}_{\ov{D}}D{}_{\ov{A}}K_{\un{C}}F_{\un{B}}{}^{\un{C}\ov{D}} + \frac{1}{2}K_{\un{C}}E_{\ov{D}}\ov{K}_{\ov{A}}F_{\un{B}}{}^{\un{C}\ov{D}} - \frac{1}{2}K_{\un{C}}E_{\ov{A}}\ov{K}_{\ov{D}}F_{\un{B}}{}^{\un{C}\ov{D}}\, \nn \\
& & + \frac{1}{2}K_{\un{C}}D_{\un{B}}\ov{K}_{\ov{A}}F^{\un{C}} + \frac{1}{2}K^{\un{C}}\ov{K}_{\ov{A}}E_{\un{C}}F_{\un{B}} - \frac{1}{2}K_{\un{B}}D_{\un{C}}\ov{K}_{\ov{A}}F^{\un{C}} + \frac{1}{2}\ov{K}_{\ov{A}}E_{\un{B}}K_{\un{C}}F^{\un{C}}\, \nn \\
& & - \frac{1}{2}\ov{K}_{\ov{A}}E_{\un{C}}K_{\un{B}}F^{\un{C}} - \frac{1}{2}\ov{K}^{\ov{E}}K_{\un{C}}F_{\ov{EDA}}F_{\un{B}}{}^{\un{C}\ov{D}} + \frac{1}{2}K^{\un{D}}\ov{K}_{\ov{E}}F_{\un{DBC}}F^{\un{C}\ov{E}}{}_{\ov{A}}\, \nn \\
& & - \frac{1}{2}K^{\un{D}}\ov{K}_{\ov{A}}F_{\un{DBC}}F^{\un{C}} - \frac{1}{2}\ov{K}^{\ov{E}}K^{\un{D}}F_{\ov{E}\un{D}}{}^{\un{C}}F_{\ov{A}\un{BC}} - 2D_{\ov{A}}D_{\un{B}}f\, ,
\eea
and
\bea
{\cal R}_{(\kappa^{2})\ov{A}\un{B}} & = & \frac{1}{2} \ov{K}^{\ov{D}}K^{\un{C}}\left(E_{\ov{D}}\ov{K}_{\ov{A}}\right)\left(E_{\un{C}}K_{\un{B}}\right) + \frac{1}{4} \ov{K}^{\ov{D}}K^{\un{C}}\ov{K}_{\ov{A}}\left(D_{\ov{D}}E_{\un{C}}K_{\un{B}}\right)\, \nn \\
& & - \frac{1}{4}K^{\un{C}}\ov{K}_{\ov{A}}K_{\un{B}}\left(E_{\un{C}}D_{\ov{D}}\ov{K}^{\ov{D}}\right) - \frac{1}{4}K^{\un{C}}\ov{K}_{\ov{A}}K^{\un{E}}E_{\un{C}}\left(\ov{K}^{\ov{D}}F_{\ov{D}\un{BE}}\right)\, \nn \\
& & - K^{\un{C}}\ov{K}_{\ov{A}}\left(E_{\un{C}}D_{\un{B}}f\right) + \left(K_{\un{B}}D_{\un{C}}\ov{K}_{\ov{A}}\right)\left(D^{\un{C}}f\right) - \left(\ov{K}_{\ov{A}}E_{\un{B}}K_{\un{C}}\right)\left(D^{\un{C}}f\right)\, \nn \\
& & + \left(\ov{K}_{\ov{A}}E_{\un{C}}K_{\un{B}}\right)\left(D^{\un{C}}f\right) + \left(K^{\un{D}}\ov{K}_{\ov{A}}F_{\un{DBC}}\right)\left(D^{\un{C}}f\right)\, .
\eea

As can be appreciated, the EOM of the generalized metric contains quadratic terms even if $f=0$, and unlike general relativity there no exist $\alpha_{1}$ and $\alpha_{2}$ such that the quadratic terms can be written as 
\bea
{{\cal R}_{(\kappa^2)}}_{\overline{A}\underline{B}} = \alpha_{1} \kappa \bar{K}_{\overline A} \bar{K}{}^{\overline C} {{\cal R}_{(\kappa)}}_{\overline{C}\underline{B}} + \alpha_{2} \kappa {K}_{\underline B} {K}^{\underline C} {{\cal R}_{(\kappa)}}_{\overline{A}\underline{C}} \, .
\nn
\eea

The previous equation shows that the equation of motion of the generalized metric cannot be linearized when the GKSA is considered. Nevertheless, upon breaking the global $O(D,D)$ invariance and using the equation of motion of $g_{\mu \nu}$ and $b_{\mu \nu}$, it is straightforward to probe that the quadratic contributions vanish when $f=0$, as showed in \cite{KL}.

\section{Higher-derivative Double Field Theory} \label{HDFT}

Higher-derivative extensions in DFT were analyzed in several works \cite{Tduality} \cite{alpha}. An iterative procedure to find an infinite tower of this kind of terms was recently given in \cite{gbdr}. In that work the authors consider an $O(D,D + K)$ multiplet $\widehat {\cal H}_{{\cal MN}}$, which is a generalized metric constrained to be an element of $O(D,D+K)$ with invariant metric $\widehat \eta_{\cal MN}$
\be
\widehat{\cal H}_{\cal MP}\ \widehat \eta^{\cal PQ}\ \widehat{\cal H}_{\cal NQ} = \widehat \eta_{\cal MN}\ , \label{HOddK}
\ee
${\cal M},{\cal N} = 0, \dots, 2d-1+K$. In this formulation, $K$ is the dimension of a gauge group $\cal{K}$ and therefore $\hat{\cal H}$ is parametrized by a generalized metric which is an $O(D,D)$ element and by a generalized constrained $O(D,D)$ vector field. 

The generalized frame $\widehat {\cal E}_{\cal M}{}^{\cal A}$ relates the generalized metric $\widehat {\cal H}_{\cal M N}$ with the flat generalized metric $\widehat {\cal H}_{\cal A B}$, and the $O(D,D+K)$ invariant metric $\widehat \eta_{\cal M N}$ with its flat version $\widehat \eta_{\cal A B}$ (which we assume to be constant) as follows
\bea
\widehat {\cal H}_{\cal M N} &=& \widehat {\cal E}_{\cal M}{}^{\cal A}\, \widehat {\cal H}_{\cal A B}\, \widehat {\cal E}_{\cal N}{}^{\cal B} \\
\widehat \eta_{\cal M N} &=& \widehat {\cal E}_{\cal M}{}^{\cal A}\,  \widehat \eta_{\cal A B}\, \widehat {\cal E}_{\cal N}{}^{\cal B} \ .
\eea

Since the idea of this formalism is to cast the $O(D,D+K)$ formulation in terms of $O(D,D)$ frame multiplets, the extended Lorentz subgroup $O(1, D - 1 + K) \to O(1,D-1) \times O(K)$ is broken such that the flat indices now split as ${\cal A} = (\underline {\cal A}\, ,\, \overline {\cal A}) = (\underline A\, ,\, \overline A \, ,\, \overline \alpha)$, and transform respectively under $O(D-1,1) \times O(1,D-1) \times O(K)$ . 

Under this splitting we have 
\bea
\widehat {\cal H}_{\cal M N} &=& - \widehat {\cal E}_{\cal M}{}^{\underline A}\, \widehat {\cal E}_{{\cal N} \underline A} + \widehat {\cal E}_{\cal M}{}^{\overline A}\, \widehat {\cal E}_{{\cal N} \overline A} +  \widehat {\cal E}_{\cal M}{}^{\overline \alpha}\,  \widehat {\cal E}_{{\cal N} \overline \alpha} \label{ODDK1}\ , \\
\widehat \eta_{\cal M N} &=& \, \ \ \widehat {\cal E}_{\cal M}{}^{\underline A}\, \widehat {\cal E}_{{\cal N} \underline A} +\widehat {\cal E}_{\cal M}{}^{\overline A}\, \widehat {\cal E}_{{\cal N} \overline A} +  \widehat {\cal E}_{\cal M}{}^{\overline \alpha}\,  \widehat {\cal E}_{{\cal N} \overline \alpha} \ , \label{ODDK}
\eea
where we use the convention that ${\cal P}_{{A} {B}}$, $\bar{\cal P}_{{A}{B}}$ and $\kappa_{\overline{\alpha} \overline{\beta}}$ raise and lower indices once we parametrize the $O(D-1,1) \times O(1, D - 1 + K)$ projectors as 
\be
\widehat {\cal P}_{{\cal A}{ \cal B}} = \left(\begin{matrix} {\cal P}_{{A}{B}} &0 & 0 \\ 0 & 0 & 0 \\ 0 & 0 & 0\end{matrix}  \right) \ , \ \ \ \
\widehat {\bar {\cal P}}_{{\cal A}{ \cal B}} = \left(\begin{matrix} 0 &0 & 0 \\ 0 & \bar {\cal P}_{{A}{B}} & 0 \\ 0 & 0 & \kappa_{\overline{\alpha} \overline{\beta}}\end{matrix}  \right) \ .
\ee

We introduce the following $O(D,D)$ multiplets
\be
{\cal H}_{MN} \ , \ \ \ \ {\cal C}_M{}^\alpha \ , \ \ \ \ d\ ,
\ee
and the $O(D,D)/O(D-1,1)\times O(1,D-1)$ frames which satisfy
\bea
\eta_{MN} &=
& {\cal H}_{MP}\ \eta^{PQ}\ {\cal H}_{NQ}   \, \nn \\ {\cal H}_{M N} &=& - {\cal E}_{M}{}^{\underline A}\, {\cal E}_{N \underline A} + \, {\cal E}_{M}{}^{\overline A}\, {\cal E}_{N \overline A} \nn \ , \\
\eta_{M N} &=& \ \ \, {\cal E}_{M}{}^{\underline A}\, {\cal E}_{N \underline A} + \, {\cal E}_{M}{}^{\overline A}\, {\cal E}_{N \overline A} 
\label{HOdd}
\eea
and we demand
\bea
\bar{\cal P}_M{}^N{\cal C}_N{}^\alpha \!&=&\! 0 \, , \quad
{\cal P}_M{}^N {\cal C}_N{}^\alpha \ = \ {\cal C}_M{}^\alpha \ . \label{Proj2}
\eea

It is straightforward to find the relation between the $O(D,D)$ multiplets and the components of the $O(D,D+K)$ multiplets. Given the following parameterization \cite{HSZ2}
\be
\widehat{\cal H}_{\cal MN} = \left(\begin{matrix}\widetilde {\cal H}_{MN} & \widetilde {\cal C}_M{}^\beta \\ (\widetilde {\cal C}^T){}^\alpha{}_N & \widetilde {\cal N}^{\alpha \beta}\end{matrix}\right)\ , \ \ \ \ \ \widehat\eta_{\cal MN} = \left(\begin{matrix}\eta_{MN} & 0 \\ 0 & \kappa^{\alpha \beta}\end{matrix}\right)\ , \label{ComponentsHoddk}
\ee
 one obtains a non-polynomial relation given by
\bea
\widetilde {\cal H}_{MN} &=& {\cal H}_{MN} + 2 {\cal C}_{M\a} \left(\kappa + {\cal C}^T \eta^{-1} {\cal C}\right)^{-1 \a \b} ({\cal C}^T)_{\b N}\ , \nn\\
\widetilde {\cal C}_{M \a} &=& 2 {\cal C}_{M \b} \left(\kappa + {\cal C}^T \eta^{-1} {\cal C}\right)^{-1 \b \gamma}\kappa_{\gamma \a}\ ,\label{ODDKtoODD}\\
\widetilde {\cal N}_{\a \b} &=& - \kappa_{\a \b} + 2 \kappa_{\a \gamma} \left(\kappa + {\cal C}^T \eta^{-1} {\cal C}\right)^{-1 \gamma \delta}\kappa_{\delta \b} \ . \nn
\eea

The tilded fields are also $O(D,D)$ multiplets, which are constrained by the requirement (\ref{HOddK}), that reads
\bea
\widetilde {\cal H}_M{}^P \widetilde {\cal H}_{NP} + \widetilde {\cal C}_M{}^\alpha \widetilde {\cal C}_{N\alpha} &=& \eta_{MN}\ ,\\
\widetilde {\cal H}_M{}^P \widetilde {\cal C}_P{}^\alpha + \widetilde {\cal C}_M{}^\beta \widetilde {\cal N}_\beta{}^\alpha &=& 0\ ,\\
\widetilde {\cal C}_{P\alpha} \widetilde {\cal C}^P{}_\beta + \widetilde {\cal N}_{\gamma\alpha} \widetilde {\cal N}^\gamma{}_{\beta} &=& \kappa_{\alpha\beta}\ .
\eea
On the other hand introducing the following definitions
\bea
\Delta_{\alpha}{}^\beta &=& \kappa_\alpha{}^\beta + {\cal C}_{M\alpha} {\cal C}^{M\beta} \label{def1}\\
\Xi_M{}^N &=& \eta_M{}^N + {\cal C}_{M \alpha} {\cal C}^{N \alpha} \  \label{def2}
\eea
we can parametrize
\bea
\widehat {\cal E}_{M}{}^{\overline A} &=& {\cal E}_M{}^{\overline A} \nn \\
\widehat {\cal E}_{M}{}^{\underline A} &=& (\Xi^{-\frac 1 2})_M{}^P\, {\cal E}_P{}^{\underline A} \nn \\
\widehat {\cal E}_{M}{}^{\overline \alpha} &=& {\cal C}_M{}^\gamma\, (\Delta^{-\frac 1 2})_\gamma{}^\beta \, e_\beta{}^{\overline \alpha} \nn \\
\widehat {\cal E}_{\alpha}{}^{\overline A} &=& 0 \nn \\
\widehat {\cal E}_{\alpha}{}^{\underline A} &=& - {\cal E}^{P \underline A}\, (\Xi^{-\frac 1 2})_P{}^Q \, {\cal C}_{Q\alpha} \nn \\
\widehat {\cal E}_{\alpha}{}^{\overline \alpha} &=& (\Delta^{-\frac 1 2})_\alpha{}^\beta \, e_\beta{}^{\overline \alpha}\ ,
\label{hatParam}
\eea
which verify (\ref{ODDK1}) and (\ref{ODDK}). Here we have introduced a constant $e_\alpha{}^{\overline \beta}$ that identifies the gauge indices $\alpha$ with $\overline \alpha$ and satisfies $e_\alpha{}^{\overline \alpha} \kappa_{\overline{\alpha} \overline{\beta}} e_\beta{}^{\overline \beta} = \kappa_{\alpha \beta}$.

\subsection{Biparametric corrections}
In this work, we explicitly implement the following identification for the gauge group ${\cal K}$, 
\bea
{\cal K} = O(1, D - 1) \subset O(1, D - 1 + K) \, ,
\eea
which is enough to include 4-derivative terms in the ordinary DFT action, as was discussed in appendix A of \cite{gbdr}. The idea is to identify the gauge degrees of freedom ${\cal C}_M{}^\alpha$ with (derivatives of) the generalized frame ${\cal E}_M{}^A$. 

Let us first begin by introducing the generators $(t_\alpha)_{\overline A}{}^{\overline B}$ that relate objects with gauge and $\bar {\cal P}$-projected adjoint Lorentz indices
\be
A_\alpha = -(t_\alpha)_{\overline B}{}^{\overline A} \, A_{\overline A}{}^{\overline B} \ , \ \ \ \ A_{\overline A}{}^{\overline B} = - A_\alpha (t^\alpha)_{\overline A}{}^{\overline B}\ . \label{identif1}
\ee
This implies that
\be
(t^\alpha)^{\overline A \overline B} \, (t_\alpha)_{\overline C \overline D} =  \delta^{\overline A}_{[\overline C} \delta^{\overline B}_{\overline D]} \ , \ \ \ \ \  (t_\alpha)_{\overline A}{}^{\overline B} \, (t_\beta)_{\overline B}{}^{\overline A} = \kappa_{\alpha \beta} \ , \label{identif2}
\ee
and
\be
[t_\alpha \, , \, t_\beta] = f_{\alpha \beta}{}^\gamma \, t_\gamma \ . \label{identif3}
\ee
We define
\be
{\cal C}_{\underline A \overline B \overline C} = - \sqrt{2} {\cal E}^{M}{}_{\underline A} {\cal C}_{M}{}^\alpha \, (t_\alpha)_{\overline B \overline C} \ 
\ee
which we identify with $\hat{{\cal F}}_{\underline A \overline B \overline C}$, as its index structure suggests. It is important to remark that this method is valid only to include four-derivative terms in the action principle, as discussed in \cite{gbdr}.

Mimicking the previous procedure, but starting with an $O(D+K,D)$ invariant theory results in an equivalent $O(D,D)$ formalism up to a $Z_2$ transformation. Therefore, the most general higher-order action principle in terms of $O(D,D)$ fields is a biparametric action with the following form,
\bea
{\cal S}  = \int d^{2D}X e^{-2d} \hat{\cal R} = \int d^{2D}X e^{-2d} \Big({\cal R} + a {\cal R}^{(-)} + b {\cal R}^{(+)} \Big)
\label{biparametric}
\eea 
where ${\cal R}^{(+)}$ is 
\bea
{\cal R}^{(+)}& = & - \frac{1}{2}\left[({\cal E}_{\underline{A}}{\cal E}_{\underline B}{\cal F}^{\underline B}{}_{\overline{CD}}) {\cal F}^{\underline{A}\overline{CD}} + ({\cal E}_{\underline{A}}{\cal E}_{\underline B}{\cal F}^{\underline A}{}_{\overline{CD}}) {\cal F}^{\underline{B}\overline{CD}}+ 2({\cal E}_{\underline A}{\cal F}_{\underline B}{}^{\overline{CD}}){\cal F}^{\underline A}{}_{\overline{CD}}{\cal F}^{\underline{B}}\right.\, \nn \\
& & + ({\cal E}_{\underline{A}}{\cal F}^{\underline{A}\overline{CD}})({\cal E}_{\underline B}{\cal F}^{\underline B}{}_{\overline{CD}}) + ({\cal E}_{\underline A} {\cal F}_{\underline B}{}^{\overline{CD}})({\cal E}^{\underline A} {\cal F}^{\underline B}{}_{\overline{CD}})+ 2({\cal E}_{\underline{A}}{\cal F}_{\underline B}){\cal F}^{\underline B}{}_{\overline{CD}}{\cal F}^{\underline{A}\overline{CD}}\, \nn \\ 
& & + ({\cal E}_{\overline{A}} {\cal F}_{\underline{B}\overline{CD}}){\cal F}_{\underline{C}}{}^{\overline{CD}}{\cal F}^{\overline{A}\underline{BC}} - ({\cal E}_{\underline{A}}{\cal F}_{\underline{B}\overline{CD}}){\cal F}_{\underline{C}}{}^{\overline{CD}}{\cal F}^{\underline{ABC}} + 2({\cal E}_{\underline A}{\cal F}^{\underline A}{}_{\overline{CD}}){\cal F}_{\underline B}{}^{\overline{CD}}{\cal F}^{\underline{B}}\, \nn \\
& &   - 4({\cal E}_{\underline A} {\cal F}_{\underline B}{}^{\overline{CD}}){\cal F}^{\underline A}{}_{\overline{CE}} {\cal F}^{\underline B\overline{E}}{}_{\overline{D}}+ \frac{4}{3}{\cal F}^{\overline{E}}{}_{{\underline A}\overline{C} }{\cal F}_{{\underline B}\overline{ED}}{\cal F}_{\underline{C}}{}^{\overline{CD}}{\cal F}^{\underline{ABC}} + {\cal F}^{\underline B}{}_{\overline{CD}} {\cal F}_{\underline A}{}^{\overline{CD}} {\cal F}_{\underline B}{\cal F}^{\underline{A}} \, \nn \\
& &  \left.+ {\cal F}_{\underline{A}}{}^{\overline{CE}}{\cal F}^{}_{\underline B\overline{E}\overline{D}}{\cal F}^{\underline A}{}_{\overline{CG}} {\cal F}^{\underline B\overline{G}\overline{D}}
 - {\cal F}_{\underline{B}}{}^{\overline{CE}}{\cal F}^{}_{\underline A\overline{E}\overline{D}}{\cal F}^{\underline A}{}_{\overline{CG}} \hat{\cal F}^{\underline B\overline{G}\overline{D}} - {\cal F}_{\overline{A}\underline{BD}}{\cal F}^{\underline{D}}{}_{\overline{CD}}{\cal F}_{\underline{C}}{}^{\overline{CD}}{\cal F}^{\overline{A}\underline{BC}}\right]\, ,\nn
 \label{Rplus}
\eea
in agreement with \cite{Tduality},
and was determined through the corrections to the extended generalized fluxes using (\ref{hatParam}),
\bea
\hat{\cal F}_{\overline A \underline{BC}} & = & {\cal F}_{\overline A \underline{BC}} +\frac{1}{2}\left({\cal E}_{\overline{A}}{\cal F}^{\overline{CD}}{}_{[\underline{B}} +{\cal F}^{\underline{E}\overline{CD}}{\cal F}_{\overline{A}\underline{E}[\underline B}\right){\cal F}_{\underline{C}]\overline{CD}}  \, , \nn \\ \hat{\cal F}_{\underline A \overline{BC}} & = & {\cal F}_{\underline A \overline{BC}} - \frac{3}{4} {\cal F}_{\underline D \overline {EF}} {\cal F}^{\overline {E F}}{}_{[\underline A} {\cal F}^{\underline D }{}_{\overline {BC}] } \, , \nn \\ \hat{\cal F}_{\underline{ABC}} & = & {\cal F}_{\underline{ABC}} +\frac{3}{2}\left(  {\cal E}_{[\underline{A}}{\cal F}^{\overline{CD}}{}_{\underline{B}} - 
\frac{1}{2}{\cal F}_{\underline D[\underline{AB}}{\cal F}^{\underline{D}\overline{CD}} - \frac{2}{3}{\cal F}^{\overline{C}}{}_{\overline{E}[\underline{A}} {\cal F}_{\underline{B}}{}^{\overline{ED}}\right) {\cal F}_{\underline{C}]\overline{CD}} \, , \nn \\ \hat{\cal F}_{\underline {A}} & = & {\cal F}_{\underline {A}} - \frac14 \left[{\cal F}^{\underline B}{}_{\overline{CD}} {\cal F}_{\underline A}{}^{\overline{CD}} {\cal F}_{\underline B} + {\cal E}_{\underline B}\left({\cal F}^{\underline B}{}_{\overline{CD}} {\cal F}_{\underline A}{}^{\overline{CD}} \right) \right] \, . 
\eea
${\cal R}^{(-)}$ coincides with ${\cal R}^{(+)}$ modulo a $Z_2$ transformation. Here $a$ and $b$ are undetermined constants and there exists an infinite amount of first-order duality invariant theories, some of them not related to String Theory \cite{HSZ}. In this work we focus in the case $a=0$ and $b=1$ in order to match with the higher-derivative heterotic supergravity after parametrization. The explicit form of the $\kappa$ terms of the 4-derivative Lagrangian is given in appendix \ref{App}.

\subsection{Equations of motion}
Using the procedure discussed in the previous section, the equations of motion of higher-derivative heterotic DFT can be recast in the following compact form,
\bea
\hat{\cal R} = {\cal R} + {\cal R}^{(+)} = 0 \, , \\
\hat{\cal R}_{\ov B \un A} = {\cal R}_{\ov B \un A} + {\cal R}^{(+)}_{\ov B \un A} = 0 \, ,
\eea
where
\bea
{\cal R}^{(+)}_{\ov{B}\un{A}} & = & -\frac{b}{4}{\cal E}_{\ov{B}}\left[{\cal E}_{\un{C}}\left({\cal F}_{\un{A}\ov{EF}}{\cal F}^{\un{C}\ov{EF}}\right) + {\cal F}_{\un{A}\ov{EF}}{\cal F}^{\un{C}\ov{EF}} {\cal F}_{\un{C}}\right] - \frac{b}{2}{\cal E}^{\un{C}}\left[\left({\cal E}_{\ov{B}}{\cal F}_{[\un{A}}{}^{\ov{EF}} + {\cal F}^{\un{D}\ov{EF}}{\cal F}_{\ov{B}\un{D}[\un{A}}\right){\cal F}_{\un{C}]\ov{EF}}\right] \nn \\ && + \frac{b}{2}\left({\cal E}_{\ov{G}}{\cal F}_{[\un{A}}{}^{\ov{EF}} + {\cal F}^{\un{D}\ov{EF}} {\cal F}_{\ov{G}\un{D}[\un{A}}\right) {\cal F}_{\un{C}]\ov{EF}} {\cal F}^{\un{C}\ov{G}}{}_{\ov{B}} + \frac{b}{4} {\cal F}_{\un{AC}}{}^{\ov{G}}{\cal F}_{\ov{BG}}{}^{\un{D}}{\cal F}_{\un{D}\ov{EF}}{\cal F}^{\un{C}\ov{EF}}\, \nn \\
& & + \frac{b}{4}{\cal E}_{\un{D}}{\cal F}_{\ov{B}\un{A}}{}^{\un{C}}{\cal F}_{\un{C}\ov{EF}}{\cal F}^{\un{D}\ov{EF}} + b {\cal E}_{[\un{A}}{\cal F}_{\un{C}]\ov{EF}}{\cal E}_{\ov{B}} {\cal F}^{\un{C}\ov{EF}} - b {\cal E}_{[\un{A}} {\cal F}_{\un{C}]\ov{EF}} {\cal F}_{\un{D}}{}^{\ov{EF}} {\cal F}_{\ov{B}}{}^{\un{CD}}\, \nn \\
& & - \frac{b}{2}{\cal E}_{\ov{B}}{\cal F}^{\un{C}\ov{EF}} {\cal F}_{\un{A}\ov{E}}{}^{\ov{G}} {\cal F}_{\un{C}\ov{GF}} + \frac{b}{2} {\cal E}_{\ov{B}} {\cal F}^{\un{C}\ov{EF}} {\cal F}_{\un{D}\ov{EF}} {\cal F}_{\un{AC}}{}^{\un{D}} - \frac{b}{2}{\cal F}_{\un{A}\ov{E}}{}^{\ov{G}}{\cal F}_{\ov{B}}{}^{\un{CD}}{\cal F}_{\un{C}}{}^{\ov{EF}}{\cal F}_{\un{D}\ov{GF}}\, \nn \\
& & + \frac{b}{2}{\cal F}_{\un{A}}{}^{\un{CD}}{\cal F}_{\ov{B}\un{CG}}{\cal F}_{\un{D}\ov{EF}}{\cal F}^{\un{G}\ov{EF}} - \frac{b}{2}\left(E_{\ov{B}}{\cal F}_{[\un{A}}{}^{\ov{EF}} + {\cal F}^{\un{D}\ov{EF}}{\cal F}_{\ov{B}\un{D}[\un{A}}\right){\cal F}_{\un{C}]\ov{EF}}{\cal F}^{\un{C}} \nn \\ && + \frac{b}{4}\left[{\cal E}_{\un{D}}\left({\cal F}_{\un{C}\ov{EF}}{\cal F}^{\un{D}\ov{EF}}\right) + {\cal F}_{\un{C}\ov{EF}}{\cal F}^{\un{D}\ov{EF}}{\cal F}_{\un{D}}\right]{\cal F}_{\ov{B}\un{A}}{}^{\un{C}}\, ,
\eea
and $\hat{\cal R}$ given in the previous section. The former equation corresponds to the first-order correction to the equation of motion of the generalized frame in the flux formalism of heterotic DFT. As far as we know, this result was not previously reported in the literature considering the flux formulation of DFT.

\subsection{Generalized Green-Schwarz transformations}

The biparametric higher-derivative DFT action (\ref{biparametric}) is invariant under generalized Lorentz transformations only if the generalized frame receives a higher-derivative correction to its Lorentz transformation,
\bea
\delta^{(1)} {\cal E}_{M \ov A} & = & - {\cal E}_{M}{}^{\un B} {\cal F}_{\un B}{}^{\ov C \ov D} {\cal E}^{\ov A}\Gamma_{\ov C \ov D} \, , \nn \\   \delta^{(1)} {\cal E}_{M \un A} & = & {\cal E}_{M}{}^{\ov{B}}{\cal F}_{\un A}{}^{\ov C \ov D} {\cal E}_{\ov{B}}\Gamma_{\ov C \ov D} \, \, , 
\label{GS}
\eea
where $\Gamma_{A B}$ was defined in (\ref{Lorentz}). Equations (\ref{GS}) mimic a Green-Schwarz mechanism, but in a DFT scenario. Using the previous expressions it is straightforward to obtain the following transformations
\bea
\delta^{(1)} {\cal F}_{\ov A \un B \un C} & = & \sqrt{2} \delta^{(1)} {\cal E}_{N \ov A} (\partial^{N}{\cal E}^{M}{}_{\un B}) {\cal E}_{M \un C} + 2 {\cal E}_{[\un B} {\cal E}^{N}{}_{\un C]} \delta^{(1)}{\cal E}_{N \ov A} \nn \\ && + 2 \sqrt{2} \delta^{(1)}{\cal E}_{N [\un B} \partial^{N}{\cal E}^{M}{}_{\un C]} {\cal E}_{M \ov A} - 2 ({\cal E}_{[\un C} \delta^{(1)}{\cal E}_{N \un B]}) {\cal E}^{N}{}_{\ov A} \, , \nn \\ \delta^{(1)} {\cal F}_{\un A \ov B \ov C} & = & \sqrt{2} \delta^{(1)} {\cal E}_{N \un A} (\partial^{N}{\cal E}^{M}{}_{\ov B}) {\cal E}_{M \ov C} + 2 {\cal E}_{[\ov B} {\cal E}^{N}{}_{\ov C]} \delta^{(1)}{\cal E}_{N \un A} \nn \\ && + 2 \sqrt{2} \delta^{(1)}{\cal E}_{N [\ov B} \partial^{N}{\cal E}^{M}{}_{\ov C]} {\cal E}_{M \un A} - 2 ({\cal E}_{[\ov C} \delta^{(1)}{\cal E}_{N \ov B]}) {\cal E}^{N}{}_{\un A} \, .
\eea
When one imposes the GKSA, the generalized background $E_{M A}$ and the generalized null vectors $K_{M}$, $\bar K_{M}$ receive a first-order Lorentz transformation coming from (\ref{GS}). This transformation can be interpreted as a generalized Green-Schwarz transformation and it must respect the constraints of the GKSA. Inspecting the null condition we need,
\bea
\delta^{(1)} (\bar K_{M}) \bar K^{M} & = & 0 \, , \quad \delta^{(1)} (K_{M}) K^{M} = 0 \, .
\label{constraint1}
\eea
Similar relations can be found inspecting the generalized geodesic equations,
\bea
\label{constraint2}
\delta^{(1)} \Big(K^{\underline A} D_{\underline A}{\bar K}^{\overline C} \Big)  =  0 \, , \quad \delta^{(1)} \Big({\bar K}^{\overline A}  {D}_{\ov A}{K}^{\underline C} \Big) & = & 0 \, .
\eea  
The previous conditions cannot be satisfied with the zeroth-order constraints for a generic solution, and therefore equations (\ref{constraint1}) and (\ref{constraint2}) can be interpreted as extra constraints of the theory. In the next part of the work we break the duality group in order to obtain the low energy effective heterotic supergravity with higher-derivative terms.

\section{Heterotic parametrization}
\label{Het}
\subsection{Parametrization}
We start by taking $D=10$ and promoting the duality group to ${\cal H}=O(9,1)_L\times O(1, 9+n)_R$ with $n=496$ in order to describe the 10-dimensional heterotic supergravity. We admit the inclusion of heterotic vectors in a duality covariant formulation as in \cite{MS2}. We split the indices as ${ M}=({}_\mu,{}^\mu,i)$ and ${ A}=(\underline{a},\overline{a},\ov{i})$. The generalized frame is parametrized in the following way,
\be
E^{M}{}_{ A}  =\left(\begin{matrix}{ E}_{\mu \underline a}&  { E}_{}^{\mu }{}_{\underline a} & E_{}^i{}_{\underline a}\\ 
E_{\mu \overline  a}& E_{}^\mu{}_{\overline  a}&E_{}^i {}_{\overline a} \\
E_{\mu\overline i} &E_{}^\mu{}_{\overline i} &E_{}^i{}_{\overline i} \end{matrix}\right) \ = \
\frac{1}{\sqrt{2}}\left(\begin{matrix}-{\tilde e}_{o\mu a}-C_{o\rho\mu} {\tilde e}_{o}^{\rho }{}_{a} &  {\tilde e}_{o}^{\mu }{}_{a} & -A_{o\rho}{}^i {\tilde e}_{o}^{\rho }{}_{{a}}\, , \\ 
\tilde{\overline e}_{o \mu a}-C_{o\rho \mu}{} \tilde{\overline e}_{o}^{\rho }{}_{{a}}& \tilde{\overline e}_{o}^\mu{}_{a}&-A_{o\rho}{}^i  \tilde{\overline e}_{o}^\rho{}_{a} \\
\sqrt{2} A_{o\mu i}e^i{}_{\overline i} &0&\sqrt{2} e^i{}_{\overline i} \end{matrix}\right)  \, ,
\label{HKparam}
\ee
where $C_{o\mu \nu}=b_{o\mu \nu} + \frac12 A_{o\mu}{}^{i} A_{o\nu i}$. The invariant projectors of DFT are parametrized in the following way
\bea
P_{\underline{ab}}=-\eta_{{ab}}\delta_{\underline a}^a\delta_{\underline b}^b, \quad \overline P_{\overline{ab}}=\eta_{{ab}}\delta_{\overline a}^a\delta_{\overline b}^b \, . \eea 
According to the previous parametrization, the generalized metric takes the following form,
\bea
{\cal H}_{M N} = \left(\begin{matrix} \tilde g_{o}^{\mu \nu} & - \tilde g_{o}^{\mu \rho} C_{o\rho \nu} & - \tilde g_{o}^{\mu \rho} A_{o \rho i} \\
- \tilde g_{o}^{\nu \rho} C_{o\rho \mu} & \tilde g_{o\mu \nu} + C_{o\rho \mu} C_{o\sigma \nu} \tilde g_{o}^{\rho \sigma} + A_{o\mu}{}^i \kappa_{ij} A_{o\nu}{}^j &
C_{o\rho \mu} \tilde g_{o}^{\rho \sigma} A_{o\sigma i} + A_{o\mu}{}^j \kappa_{ji} \\
- \tilde g_{o}^{\nu \rho} A_{o\rho i} & C_{o\rho \nu} \tilde g_{o}^{\rho \sigma} A_{o\sigma i} +  A_{o\nu}{}^j \kappa_{ij} & \kappa_{ij} + A_{o\rho i} \tilde g_{o}^{\rho \sigma} A_{o\sigma j}\end{matrix}\right) \ .
\label{Gmetric}
\eea
On the other hand $K_{M}$ and ${\bar K}_{M}$ can be parametrized as
\bea
K_{M} = \, \frac{1}{\sqrt{2}} \left( \begin{matrix} l^{\mu} \\ - l_{\mu} - C_{o \rho \mu} l^{\rho } \\ - A_{oi \rho} {l}^{\rho} \end{matrix} \right) \, , \quad
\bar{K}_{M} = \, \frac{1}{\sqrt{2}} \left( \begin{matrix} {\bar l}^{\mu} \\  {\bar l}_{\mu} - C_{o \rho \mu} {\bar l}^{\rho} \\- A_{oi \rho} {\bar l}^{\rho} \end{matrix} \right) \, . 
\eea

We impose the standard gauge fixing for the double Lorentz group,
\bea
\tilde e_{o\mu a} \eta^{ab} \tilde e_{o\nu b} = \tilde{\overline e}_{o\mu a} \eta^{ab} \tilde{\overline e}_{o\nu b} = \tilde{g}_{\mu \nu} \, , 
\label{gf}
\eea
with $\eta_{ab}$ the ten dimensional flat metric, $a,b=0,\dots, 9$. Finally, the parametrization of the generalized dilaton is,
\bea
 e^{-2d} = \sqrt{\tilde g} e^{-2 \tilde \phi} \, .
\eea

The previous parametrization reproduce the low energy heterotic supergravity with higher-derivative terms. While $\tilde{g}_{\mu \nu}$ and $b_{\mu \nu}$ are consistently perturbed by a pair of null vectors $l$ and $\bar{l}$ as in (\ref{metricp}) and (\ref{bp}), the perturbations of the gauge field $A_{o \mu i}$ are suppressed by the $O(10,10+n)$ invariance.

\subsection{Field redefinitions}

One of the most interesting aspects of higher-derivative DFT is the need of field redefinitions to match with standard heterotic supergravity using (\ref{HKparam}). 
It is straightforward to show that ${\tilde g}^{\mu \nu}$ transforms under Lorentz transformations as,
\bea
\delta_{\Lambda} {\tilde g}^{\mu \nu} = \delta_{\Lambda}(\tilde g_{o}^{\mu \nu} + \kappa l^{(\mu} \bar{l}^{\nu)}) = - \Omega_{(\mu}{}^{a b} \partial_{\nu)} \Lambda_{a b} \, .  
\eea
We stress that this field redefinition is independent of the GKSA and therefore it is mandatory to  consider an exact metric and dilaton redefinition of the form \cite{Tduality}, 
\bea
\label{metricredef}
\tilde{g}_{\mu \nu} & = & g_{\mu \nu} - \frac{1}{2}  \Omega^{(-)}_{\mu ab} \Omega^{(-)}_{\nu}{}^{ab} \, , \\
\sqrt{-\tilde g}e^{-2\tilde{\phi}} & = & \sqrt{-g}e^{-2{\phi}} \, ,
\eea
and hence $
\delta_{\Lambda} g_{\mu \nu} = 0$
as in (\ref{transf1}). The previous field redefinitions can be imposed at the level of perturbative double field theory and, in particular, when one considers the GKSA. Parametrizing the generalized frame perturbations we find (before imposing the gauge fixing),
\bea
\tilde{\bar e}_{\mu}{}^{\overline a} = \tilde{\bar e}_{o \mu}{}^{\overline a} - \frac12 \wt{\kappa} \bar l_{\mu} l_{\nu} \tilde{\bar e}_{o}^{\nu \overline a} \, , \\
\tilde{ e}_{\mu}{}^{\underline a} = \tilde{e}_{o \mu}{}^{\underline a} - \frac12 \wt{\kappa} l_{\mu} \bar l_{\nu} \tilde{e}_{o}^{\nu \underline a} \, , \\
\tilde{\bar e}^{\mu \overline a} = \tilde{\bar e}_{o }{}^{\mu \overline a} + \frac12 \kappa l^{\mu} \bar l^{\nu} \tilde{\bar e}_{o \nu}{}^{\overline a} \, ,  \\
\tilde{e}^{\mu \underline a} = \tilde{e}_{o }{}^{\mu \overline a} + \frac12 \kappa \bar l^{\mu}  l^{\nu} \tilde{e}_{o \nu}{}^{\underline a} \, ,
\eea
where 
\bea
\wt{\kappa}=\frac{2\kappa}{2+\kappa (l\cdot\bar l)}.
\eea
In the limit $\kappa^2 \sim 0$ we can identify $\tilde {\bar e}_{\mu}{}^{\ov a} \rightarrow \tilde e_{\mu}{}^{a}$ (and $\tilde {\bar e}^{\mu \ov a} \rightarrow \tilde{e}^{\mu a}$). Moreover, since the field redefinition (\ref{metricredef}) contains two explicit derivatives we can construct the torsionful connection using $e_{\mu a}$ instead of $\tilde e_{\mu a}$. The former is perturbed as 
\bea
e_{\mu}{}^{a} = e_{o \mu}{}^{a} - \frac12 \kappa \bar l_{\mu} l_{\nu} e_{o}^{\nu \overline a}
\eea
and the torsionful connection can be easily constructed considering the perturbations of the spin connection
\be
w_{\mu a b} = w_{o \mu a b} - \frac{1}{2}\kappa\left[\nabla_{o\mu}\left(l_{\nu}\bar{l}_{\sigma}\right) + \nabla_{o\nu}\left(l_{\sigma}\bar{l}_{\mu}\right) - \nabla_{o\sigma}\left(l_{\mu}\bar{l}_{\nu}\right)\right]e_{o}^{\nu}{}_{[a}e_{o}^{\sigma}{}_{b]} + {\cal O}(\kappa^2)\, , 
\ee
and the 3-form
\bea
H_{\mu \nu \rho} = H_{o \mu \nu \rho} - 3\kappa\nabla_{o[\mu}\left(l_{\nu}\bar{l}_{\rho]}\right) + {\cal O}(\kappa^2) \, . 
\eea
In this case the gauge fixing implies $\tilde e_{o \mu}{}^{\un a} \delta_{\un a}{}^{a} = e_{o \mu}{}^{a} - \tilde \kappa \bar l_{[\mu} l_{\nu]} e_{o}{}^{\nu a}$, where the left background vielbein is related to right one through $l$ and $\bar l$ terms.  

In what follows we use the present formulation to find higher-derivative corrections in the context of heterotic supergravity considering the GKSA. We restrict our study to the leading order in $\kappa$ terms in order to be compatible with the gauge fixing here presented.

\section{Classical Double Copy}
\label{CDC}
The double-copy structure of perturbative gravity originates from string theory in the so called KLT formalism, where one identifies universal relations between open- and closed-string
tree-level amplitudes \cite{KLT}. The KLT relations have been later on reformulated in a field-theory
framework by Bern, Carrasco and Johansson (BCJ) noticing a hidden symmetry of gauge-theory amplitudes which is a duality between color and kinematics \cite{BCJ}. In heterotic supergravity, the identification of the null vectors with a pair of $U(1)$ gauge fields reproduce a pair of Maxwell-like equations to describe the dynamics of the system, as showed in \cite{HetKL}. In this part of the work we inspect higher-derivative corrections to these equations.  

\subsection{Double null vector ansatz}
\label{DC2}
Higher-derivative terms can be easily incorporated in the classical double copy prescription of the low energy limit of heterotic string theory. We assume that the geometry admits one Killing vector $\xi^{\mu}$ such that the Lie derivative ${L}_{\xi}$ acting on an exact field vanishes,
\bea
L_{\xi} T_{\mu_{1} \mu_{2} \mu_{3} \dots} =0 \, 
\label{Lie}
\eea
where $T_{\mu_{1} \mu_{2} \mu_{3} \dots}$ is an arbitrary tensor. Moreover we choose a coordinate system where $\xi^{\mu}$ is covariantly constant, \textit{i.e.},
\bea
\nabla_{o\mu}\xi_{\nu} = \nabla_{o[\mu}\xi_{\nu]} = 0 \, ,
\eea
and then condition (\ref{Lie}) is
\bea
\xi^{\mu} \nabla_{o \mu} T_{\mu_{1} \mu_{2} \mu_{3} \dots} = 0 \, .
\eea
We normalize the null vectors to satisfy,
\bea
\xi_{\mu} \eta^{\mu \nu} l_{\mu} = \xi_{\mu} \eta^{\mu \nu} \bar l_{\mu} = 1 \, .
\eea

In order to obtain the leading order terms with up to four derivatives in the single and zeroth copy, we perturbe the gravity contributions to (\ref{Deltag}) by considering $A_{o\mu i}=0$, $\phi=\phi_{o}=\textrm{const.}$ and $\hat H_{\mu \nu \rho}=\hat H_{o \mu \nu \rho}=0$. Then we contract the equation of motion for the metric tensor with $\xi^{\mu}$ and $\xi^{\mu} \xi^{\nu}$,
\bea
\label{DCg0}
\xi^{\mu} \Delta^{(1)} g_{\mu \nu} & = &  \, - \frac {1} 2 \xi^{\mu} g^{\sigma \tau } R_{\sigma \mu a b}  R_{\tau \nu}{}^{a b} \, , \\ \xi^{\mu} \xi^{\nu} \Delta^{(1)} g_{\mu \nu} & = & - \frac {1} 2 \xi^{\mu} \xi^{\nu} g^{\sigma \tau } R_{\sigma \mu a b}  R_{\tau \nu}{}^{a b} \, ,
\label{DCg}
\eea
to find the single copy and the zeroth copy, respectively. We perturb around a generic background,
\bea
g^{\mu \nu} & = & g_{o}^{\mu \nu} + \kappa \varphi l^{(\mu} \bar{l}^{\nu)} \, , \nn \\
g_{\mu \nu} & = & g_{o\mu \nu} - \kappa \varphi l^{(\mu} \bar{l}^{\nu)} \, ,
\eea
keeping only $\kappa$ terms to be compatible with the previous section, and we include the scalar function $\varphi$ in the ansatz.

After imposing the previous ansatz the connection takes the following form 
\bea
\Gamma^{\sigma}_{\mu\nu} & = & \Gamma^{\sigma}_{o\mu\nu} - \frac{1}{2}{\kappa}g_{o}^{\sigma\rho}\left[\nabla_{o\mu}\left(\varphi \bar{l}_{(\nu}l_{\rho)}\right) + \nabla_{o\nu}\left(\varphi \bar{l}_{(\mu}l_{\rho)}\right) - \nabla_{o\rho}\left(\varphi \bar{l}_{(\mu}l_{\nu)}\right)\right] \, .
\eea
The Riemann tensor for this configuration can be written in the following covariant way,
\bea
R^{\sigma}{}_{\lambda\mu\nu} & = & R_{o}^{\sigma}{}_{\lambda\mu\nu} - \frac{1}{2}\kappa g_{o}^{\sigma\rho}\nabla_{o\mu}\left[\nabla_{o\nu}\left(\varphi \bar{l}_{(\lambda}l_{\rho)}\right) + \nabla_{o\lambda}\left(\varphi \bar{l}_{(\nu}l_{\rho)}\right) - \nabla_{o\rho}\left(\varphi \bar{l}_{(\nu}l_{\lambda)}\right)\right]\, \nn \\
& & + \frac{1}{2}\kappa g_{o}^{\sigma\rho}\nabla_{o\nu}\left[\nabla_{o\mu}\left(\varphi \bar{l}_{(\lambda}l_{\rho)}\right) + \nabla_{o\lambda}\left(\varphi \bar{l}_{(\mu}l_{\rho)}\right) - \nabla_{o\rho}\left(\varphi \bar{l}_{(\mu}l_{\lambda)}\right)\right] + \mathcal{O}(\kappa^2)\, . \nn
\eea
The leading order contributions to the equation of motion of the metric tensor were studied in \cite{HetKL} and match with Maxwell-like equations after identifing $\varphi l_{\mu} = A_{\mu}$ and $\varphi \bar{l}_{\mu}={\bar A}_{\mu}$, where $A_{\mu}$ and ${\bar A}_{\mu}$ are a pair of $ U(1)$ gauge vectors,
\bea
\frac{\kappa}{4} \nabla_{o}^{\mu} F_{\mu \nu i} & = & 0 \, , \\
\frac{\kappa}{4} \nabla_{o}^{\mu} \bar F_{\mu \nu i} & = & 0 \, .
\eea
The curvatures for the abelian gauge fields are $F_{\mu \nu i} = 2 \partial_{[\mu} A_{\nu]i}$ and $\bar F_{\mu \nu i} = 2 \partial_{[\mu} \bar A_{\nu]i}$. The first correction to these equations are $\kappa$ terms that come from the linear perturbation of the Riemann tensor in (\ref{DCg}). Explicitly,
\bea
\frac{\kappa}{4} \left[ \nabla_{o}^{\mu} F_{\mu \nu} + (\nabla_{o\rho}F_{\mu \sigma}) R_{o\nu}{}^{\mu\rho\sigma} \right] & = & 0 \, , \\
\frac{\kappa}{4} \left[ \nabla_{o}^{\mu} \bar F_{\mu \nu} + (\nabla_{o\rho}\bar F_{\mu \sigma}) R_{o\nu}{}^{\mu\rho\sigma} \right] & = & 0 \, ,
\eea
where we have used
\be
\label{kill_riem}
[\nabla_{o\rho},\nabla_{o\sigma}]\xi^{\mu} = R_{o}^{\mu}{}_{\lambda\rho\sigma}\xi^{\lambda} = 0\, .
\ee
The zeroth copy dynamics does no receive a higher-derivative correction in this approximation,
\bea
\frac{\kappa}{4} \nabla_{o}^{\mu} \nabla_{o\mu}\varphi = 0 \, .
\eea

Finally we mention that the contributions found in this paper are consistent with the KLT relation but they do not satisfy the color-kinematics duality. The gauge contributions that satisfy this duality are \cite{Garozzo}
\bea
L_{\rm{open}} = \frac{1}{4} F_{\mu \nu i} F^{\mu \nu i} + \frac{2}{3} F_{\mu \nu} F^{\nu \lambda} F_{\lambda}{}^{\mu} + \mathcal{O}(F^4) \, , 
\label{open}
\eea
where $L_{\rm{open}}$ is the effective open string Lagrangian. The second contribution in (\ref{open}) is a ${\cal O}(\alpha')$ contribution that requires non-abelian contributions from the structure constants of the heterotic gauge group. We left the study of these color-kinematics terms for future work. 

\section{Conclusions}
\label{Con}

We study the heterotic formulation of DFT when higher-derivative terms are included, and the field content is perturbed with the GKSA. We start by adapting the GKSA to the flux formulation of DFT. Then we compute a higher-derivative extension for DFT considering multiplets of $O(D,D+K)$ and we choose the free parameters of the formalism to match with the heterotic case. At this stage we review the four-derivative corrections to the action principle of DFT and we compute the full first order equations of motion. Then we impose the GKSA to compute the leading-order contributions to the action principle and we study the first order symmetry corrections in this framework. The double Lorentz symmetry is deformed by a generalized Green-Schwarz transformation that must respect the constraints of the GKSA and these are new conditions for generic double backgrounds. 

Upon parametrization, we reproduce the low energy heterotic supergravity with higher-derivative terms. Higher-derivative field redefinitions are required to match with the standard transformation rules. The gravitational field content, $g_{\mu \nu}$ and $b_{\mu \nu}$, is consistently perturbed by a pair of null vectors $l$ and $\bar{l}$. Interestingly enough, the perturbations of the gauge field $A_{\mu i}$ are suppressed by the $O(10,10+n)$ invariance, using the flux formulation of DFT. This last point indicates a tension between the generalized metric formalism and the generalized frame formalism upon parametrization. Moreover the generalized frame formulation requires to impose gauge fixing to relate the vielbeins needed to construct the torsionful spin connection. Here we solve this issue considering $\kappa^2 \sim 0$. 

As an application we study higher-derivative contributions to the classical double copy. We focus in the single and zeroth copy coming from the $\rm{Riem}^2$ starting from a generic background. In this scenario we obtain four-derivative corrections to the $\kappa$ Maxwell-like equations previously discussed in \cite{HetKL} in agreement with the KLT relation.  

\subsection*{Acknowledgements}
We sincerely thank K. Lee and K. Cho for exceedingly interesting remarks and comments. Support by CONICET is also gratefully acknowledged.

\appendix

\section{Corrections to the DFT action}
\label{App}
The four-derivative contributions to the DFT Lagrangian when the GKSA is imposed are given by
\bea
{\cal R} = R + b \kappa (T_{0} + T_{1} + T_{2} + T_{3}) + b {\cal O}(\kappa^2) \, ,
\eea
where
{\footnotesize
\bea
T_{0} & = & K_{{\un A}} \ov K_{{\ov A}} ( - \frac12 F^{{\ov A}} F^{{\un B}} F_{{\un B}}{}^{{\ov B} {\ov C}} F^{{\un A}}{}_{{\ov B} {\ov C}} - 2 F^{{\ov B} {\un B} {\un A}} F_{{\un B} {\un C} {\un D}} F^{{\un C}}{}_{{\ov B}}{}^{{\ov C}} F^{{\un D}}{}_{{\ov C}}{}^{{\ov A}}  - \frac12 F^{{\un B}} F^{{\un A}} F^{{\ov B} {\ov C} {\ov A}} F_{{\un B} {\ov B} {\ov C}} \nn \\ && - F^{{\ov B} {\ov C} {\ov A}} F_{{\un B} {\ov D}}{}^{{\ov E}} F^{{\un A}}{}_{{\ov C} {\ov E}} F^{{\un B}}{}_{{\ov B}}{}^{{\ov D}} - 2 F^{{\ov B} {\un B} {\un A}} F_{{\un C} {\ov C}}{}^{{\ov A}} F_{{\un B}}{}^{{\ov D} {\ov C}} F^{{\un C}}{}_{{\ov B} {\ov D}} + F^{{\ov B} {\ov C} {\ov A}} F_{{\un B} {\ov C}}{}^{{\ov D}} F^{{\un A}}{}_{{\ov E} {\ov D}} F^{{\un B}}{}_{{\ov B}}{}^{{\ov E}} \nn \\ && - \frac12 F^{{\un B}} F^{{\ov B} {\ov C} {\ov D}} F_{{\un B} {\ov B} {\ov C}} F^{{\un A}}{}_{{\ov D}}{}^{{\ov A}} - F^{{\ov B} {\un B} {\un A}} F_{{\un C} {\ov C} {\ov D}} F_{{\un B}}{}^{{\ov C} {\ov A}} F^{{\un C}}{}_{{\ov B}}{}^{{\ov D}} + F^{{\ov B} {\un B} {\un A}} F_{{\un B} {\ov B}}{}^{{\ov C}} F_{{\un C} {\ov D}}{}^{{\ov A}} F^{{\un C}}{}_{{\ov C}}{}^{{\ov D}} \nn \\ && - \frac12 F^{{\ov B} {\un B} {\un A}} F_{{\un C} {\ov C} {\ov D}} F_{{\un B}}{}^{{\ov C} {\ov D}} F^{{\un C}}{}_{{\ov B}}{}^{{\ov A}} + \frac12 F_{{\ov B} {\un B}}{}^{{\un A}} F^{{\ov C} {\ov D} {\ov A}} F^{{\ov B} {\un B} {\un C}} F_{{\un C} {\ov C} {\ov D}} + F^{{\ov A} {\un B} {\un C}} F_{{\un C} {\ov B}}{}^{{\ov C}} F_{{\un B}}{}^{{\ov B} {\ov D}} F^{{\un A}}{}_{{\ov D} {\ov C}} \nn \\ && - \frac12 F^{{\un B}} F^{{\ov A} {\un C} {\un A}} F_{{\un C} {\ov B} {\ov C}} F_{{\un B}}{}^{{\ov B} {\ov C}} - F^{{\un B}} F^{{\un C}} F^{{\ov B}}{}_{{\un B}}{}^{{\un A}} F_{{\un C} {\ov B}}{}^{{\ov A}} - \frac12 F^{{\ov B} {\un B} {\un C}} F_{{\un B} {\ov B}}{}^{{\ov A}} F_{{\un C}}{}^{{\ov C} {\ov D}} F^{{\un A}}{}_{{\ov C} {\ov D}} \nn \\ && + F^{{\ov B} {\ov C} {\ov A}} F_{{\un B} {\un C}}{}^{{\un A}} F^{{\un C}}{}_{{\ov C} {\ov D}} F^{{\un B}}{}_{{\ov B}}{}^{{\ov D}} + \frac12 F^{{\ov A} {\un B} {\un C}} F_{{\un B}}{}^{{\ov B} {\ov C}} F_{{\un C} {\un D}}^{{\un A}} F^{{\un D}}{}_{{\ov B} {\ov C}} + F_{{\ov B} {\un B}}{}^{{\un C}} F^{{\ov C}}{}_{{\un D}}{}^{{\un A}} F^{{\ov B} {\un B} {\un D}} F_{{\un C} {\ov C}}{}^{{\ov A}}) \, ,
\eea}
\footnotesize{
\bea
T_{1} & = & - 2  E^{\un A}{f} F^{\un B} F_{\un B \ov A \ov B} F_{\un A}{}^{\ov A \ov B} +  K_{\un A} {\ov K}_{\ov A} ( - \frac12 E^{\ov A}{F^{\un A}{}_{\ov B \ov C}} F^{\un B} F_{\un B}{}^{\ov B \ov C} \nn  + \frac12 E^{\un A}{F_{\un B \ov B \ov C}} F^{\un B} F^{\ov A \ov B \ov C} \\ && - \frac12  E^{\un A}{F_{\un B}} F^{\ov B \ov C \ov A} F^{\un B}{}_{\ov B \ov C}  - E^{\un A}{F_{\un B \ov B \ov C}} F^{\ov A \un C \un B} F_{\un C}{}^{\ov B \ov C} - \frac12 E^{\un A}{F_{\ov B \ov C}{}^{\ov A}} F^{\un B} F_{\un B}{}^{\ov B \ov C} \nn \\ && - \frac12 E^{\ov B}{F^{\un A}{}_{\ov C \ov D}} F_{\un B}{}^{\ov C \ov D} F^{\un B}{}_{\ov B}{}^{\ov A} + \frac12 E^{\un A}{F_{\un B \ov B \ov C}} F^{\ov D \ov A \ov B} F^{\un B}{}_{\ov D}{}^{\ov C} + \frac12 E^{\ov A}{F_{\un B \ov B \ov C}} F^{\un B} F^{\un A \ov B \ov C} \nn \\ && - \frac12 E^{\un B}{F^{\ov A}} F_{\un B}{}^{\ov B \ov C} F^{\un A}{}_{\ov B \ov C} - \frac12 E^{\ov A}{F_{\un B}} F^{\un B}{}_{\ov B \ov C} F^{\un A \ov B \ov C} + \frac12 E^{\ov A}{F_{\un B \ov B \ov C}} F^{\un B}{}_{\ov D}{}^{\ov B} F^{\un A \ov D \ov C} \nn \\ && + E^{\un B}{F_{\ov B \un B}{}^{\un A}} F^{\un C} F_{\un C}{}^{\ov A \ov B} + \frac12 E^{\un B}{F^{\un A}{}_{\ov B \ov C}} F^{\ov A} F_{\un B}{}^{\ov B \ov C}  + \frac12 E^{\un B}{F_{\ov B \ov C}{}^{\ov A}} F^{\un A} F_{\un B}{}^{\ov B \ov C} \nn \\ && - \frac12  E^{\un B}{F^{\un A}} F^{\ov B \ov C \ov A} F_{\un B \ov B \ov C} - E^{\ov B}{F_{\ov C \un B}{}^{\un A}} F_{\ov B}{}^{\un C \un B} F_{\un C}{}^{\ov A \ov C} - \frac12 E^{\un B}{F^{\un A}{}_{\ov B \ov C}} F^{\ov A}{}_{\un B}{}^{\un C} F_{\un C}{}^{\ov B \ov C} \nn \\ && - \frac12 E^{\un B}{F_{\un B \ov B \ov C}} F^{\ov A} F^{\un A \ov B \ov C} - \frac12 E^{\un B}{F_{\un B \ov B \ov C}} F^{\un A} F^{\ov A \ov B \ov C}  - \frac12 E^{\un B}{F_{\ov B \ov C}{}^{\ov A}} F_{\un B}{}^{\ov D \ov B} F^{\un A}{}_{\ov D}{}^{\ov C} \nn \\ && + \frac12 E^{\ov B}{F_{\un B \ov C \ov D}} F^{\un B}{}_{\ov B}{}^{\ov A} F^{\un A \ov C \ov D} + \frac12 E^{\ov B}{F_{\un B \ov C \ov D}} F_{\ov B}{}^{\un A \un B} F^{\ov A \ov C \ov D} + \frac12 E^{\ov B}{F_{\ov C \ov D}{}^{\ov A}} F_{\ov B}{}^{\un B \un A} F_{\un B}{}^{\ov C \ov D} \nn \\ && - \frac12 E^{\un B}{F^{\un A}{}_{\ov B \ov C}} F^{\ov D \ov A \ov B} F_{\un B \ov D}{}^{\ov C} - \frac12 E^{\un B}{F^{\ov A}{}_{\un C}{}^{\un A}} F_{\un B}{}^{\ov B \ov C} F^{\un C}{}_{\ov B \ov C}  +  E^{\ov A}{F_{\un B \ov B \ov C}} F_{\un C}{}^{\un A \un B} F^{\un C \ov B \ov C} \nn \\ &&
+ E^{\un B}{F_{\un C \ov B}{}^{\ov A}} F^{\un C} F^{\ov B}{}_{\un B}{}^{\un A}  - E^{\un B}{F_{\ov B \un C}{}^{\un A}} F^{\un C} F_{\un B}{}^{\ov A \ov B} - E^{\un B}{F_{\un C}} F^{\ov B}{}_{\un B}{}^{\un A} F^{\un C}{}_{\ov B}{}^{\ov A} \nn \\ && + E^{\un B}{F_{\un C}} F^{\ov B \un A \un C} F_{\un B \ov B}{}^{\ov A} + E^{\un B}{F_{\ov B \un C}{}^{\un A}} F_{\un B \un D}{}^{\un C} F^{\un D \ov A \ov B}  - E^{\un B}{F_{\un B \ov B}{}^{\ov A}} F^{\un C} F^{\ov B}{}_{\un C}{}^{\un A} \nn \\ && + \frac12 E^{\un B}{F_{\un C \ov B}{}^{\ov A}} F^{\ov C \un A \un C} F_{\un B \ov C}{}^{\ov B}  + \frac12 E^{\un B}{F_{\un C \ov B}{}^{\ov A}} F^{\ov C}{}_{\un B}{}^{\un A} F^{\un C}{}_{\ov C}{}^{\ov B} - \frac12 E^{\un B}{F_{\un C \ov B \ov C}} F^{\ov A \un A \un C} F_{\un B}{}^{\ov B \ov C} \nn \\ && + E^{\ov B}{F_{\un B \ov C}{}^{\ov A}} F_{\ov B \un C}{}^{\un B} F^{\ov C \un C \un A} + \frac12 E^{\un B}{F_{\un C \ov B \ov C}} F^{\ov A}{}_{\un B}{}^{\un C} F^{\un A \ov B \ov C}  - \frac12 E^{\un B}{F_{\un C \ov B \ov C}} F^{\ov A \ov B \ov C} F_{\un B}{}^{\un A \un C} \nn \\ && - \frac12 E^{\un B}{F_{\ov B \ov C}{}^{\ov A}} F_{\un B \un C}{}^{\un A} F^{\un C \ov B \ov C}  - \frac12 E^{\un B}{F_{\un B \ov B \ov C}} F^{\ov A \un C \un A} F_{\un C}{}^{\ov B \ov C} + \frac12 E^{\un B}{F_{\un C \ov B \ov C}} F^{\ov B \un A \un C} F_{\un B}{}^{\ov A \ov C} \nn \\ &&
+ \frac12 E^{\un B}{F_{\un C \ov B \ov C}} F^{\ov B}{}_{\un B}{}^{\un A} F^{\un C \ov A \ov C} - \frac12 E^{\un B}{F_{\ov B \un C}{}^{\un A}} F_{\un B}{}^{\ov C \ov B} F^{\un C}{}_{\ov C}{}^{\ov A}  + \frac12 E^{\un B}{F_{\ov B \un C}{}^{\un A}} F_{\un B}{}^{\ov C \ov A} F^{\un C}{}_{\ov C}{}^{\ov B} \nn \\ && - E^{\un B}{F_{\un C \ov B}{}^{\ov A}} F^{\ov B \un D \un A} F_{\un B \un D}{}^{\un C}) 
- \frac12  K_{a} E^{\ov A}{{\ov K}_{\ov A}} F^{\un B} F_{\un B}{}^{\ov B \ov C} F^{\un A}{}_{\ov B \ov C} - \frac12  {\ov K}_{\ov A} E^{\un A}{K_{\un A}} F^{\un B} F^{\ov B \ov C \ov A} F_{\un B \ov B \ov C} \nn \\ && +  K_{\un A} E^{\ov A}{\ov K_{\ov B}} (\frac12 F_{\ov A}{}^{\un B \un A} F^{\ov C \ov D \ov B} F_{\un B \ov C \ov D} + F^{\un B} F^{\un A} F_{\un B \ov A}{}^{\ov B}  - F_{\un B \ov C}{}^{\ov D} F^{\un B}{}_{\ov A}{}^{\ov C} F^{\un A}{}_{\ov D}{}^{\ov B} \nn \\ && + F_{\un B \ov C}{}^{\ov B} F^{\un A}{}_{\ov A \ov D} F^{\un B \ov C \ov D} \nn  + 2 F_{\un B}{}^{\ov C \ov B} F^{\un A}{}_{\ov D \ov C} F^{\un B}{}_{\ov A}{}^{\ov D} - F_{\ov C \un B}{}^{\un A} F^{\ov C \un B \un C} F_{\un C \ov A}{}^{\ov B} \nn \\ && - F_{\ov A}{}^{\un B \un C} F^{\ov C}{}_{\un B}{}^{\un A} F_{\un C \ov C}{}^{\ov B} + 2 F_{\un B \un C}{}^{\un A} F^{\un B}{}_{\ov A}^{\ov C} F^{\un C}{}_{\ov C}{}^{\ov B})  +  K_{\un A} E^{\un B}{{\ov K}_{\ov A}} (\frac12 F^{\un A} F^{\ov B \ov C \ov A} F_{\un B \ov B \ov C} \nn
\eea
\footnotesize{
\bea
 && - \frac12 F^{\ov A} F_{\un B}{}^{\ov B \ov C} F^{\un A}{}_{\ov B \ov C}  - \frac12 F^{\ov B \ov C \ov A} F_{\un B \ov B}{}^{\ov D} F^{\un A}{}_{\ov C \ov D} + F^{\un C} F^{\ov B}{}_{\un C}{}^{\un A} F_{\un B \ov B}{}^{\ov A} \nn \\ && + F^{\ov B \un C \un A} F_{\un B \un C \un D} F^{\un D}{}_{\ov B}{}^{\ov A} - \frac12 F^{\ov B \ov C \ov A} F_{\un B \un C}{}^{\un A} F^{\un C}{}_{\ov B \ov C}  - F^{\un C} F^{\ov B}{}_{\un B}{}^{\un A} F_{\un C \ov B}{}^{\ov A} \nn \\ && + \frac12 F^{\ov B \un C \un A} F_{\un B \ov B}{}^{\ov C} F_{\un C \ov C}{}^{\ov A}  - \frac12 F^{\ov B \un C \un A} F_{\un C \ov B \ov C} F_{\un B}{}^{\ov C \ov A} + \frac12 F^{\ov A}{}_{\un B}{}^{\un C} F_{\un C}{}^{\ov B \ov C} F^{\un A}{}_{\ov B \ov C}) \nn \\ && +  {\ov K}_{\ov A} E^{\ov B}{K_{\un A}} (F^{\un A} F^{\un B} F_{\un B \ov B}{}^{\ov A} - F_{\ov B}{}^{\un B \un C} F^{\ov C}{}_{\un B}{}^{a} F_{\un C \ov C}{}^{\ov A} + 2 F_{\un B \un C}{}^{\un A} F^{\un B}{}_{\ov B}{}^{\ov C} F^{\un C}{}_{\ov C}{}^{\ov A} \nn \\ && + F_{\un B \ov C}{}^{\ov A} F^{\un A}{}_{\ov B \ov D} F^{\un B \ov C \ov D} + 2 F_{\un B}{}^{\ov C \ov A} F^{\un A}{}_{\ov D \ov C} F^{\un B}{}_{\ov B}{}^{\ov D} - F_{\ov C \un B}{}^{\un A} F^{\ov C \un B \un C} F_{\un C \ov B}{}^{\ov A} \nn \\ && + \frac12 F_{\ov B}{}^{\un B \un A} F^{\ov C \ov D \ov A} F_{\un B \ov C \ov D} - F_{\un B \ov C}{}^{\ov D} F^{\un A}{}_{\ov D}{}^{\ov A} F^{\un B}{}_{\ov B}{}^{\ov C}) +  {\ov K}_{\ov A} E^{\un A}{K_{\un B}} ( - \frac12 F^{\ov A} F_{\un A}{}^{\ov B \ov C} F^{\un B}{}_{\ov B \ov C} \nn \\ && + \frac12 F^{\ov A}{}_{\un A}{}^{\un C} F_{\un C}{}^{\ov B \ov C} F^{\un B}{}_{\ov B \ov C}  + F^{\un C} F^{\ov B}{}_{\un C}{}^{\un B} F_{\un A \ov B}{}^{\ov A} + F^{\ov B \un C \un B} F_{\un A \un C \un D} F^{\un D}{}_{\ov B}{}^{\ov A} \nn \\ && - F^{\un C} F^{\ov B}{}_{\un A}{}^{\un B} F_{\un C \ov B}{}^{\ov A} + \frac12 F^{\un B} F^{\ov B \ov C \ov A} F_{\un A \ov B \ov C}  - \frac12 F^{\ov B \ov C \ov A} F_{\un A \un C}{}^{\un B} F^{\un C}{}_{\ov B \ov C} \nn \\ && + \frac12 F^{\ov B \un C \un B} F_{\un A \ov B}{}^{\ov C} F_{\un C \ov C}{}^{\ov A}  - \frac12 F^{\ov B \un C \un B} F_{\un C \ov B \ov C} F_{\un A}{}^{\ov C \ov A} - \frac12 F^{\ov B \ov C \ov A} F_{\un A \ov B}{}^{\ov D} F^{\un B}{}_{\ov C \ov D})
\eea}
\footnotesize{
\bea
T_2 = && 2 E^{\un A}{F_{\un B \ov A \ov B}} E^{\un B}{f} F_{\un A}{}^{\ov A \ov B} - 2 \sqrt{2} E^{\un A}{E^{M}{}_{\un B}} F_{\un A}{}^{\ov A \ov B} F^{\un B}{}_{\ov A \ov B} \partial_{M}{f}  - 4 E^{M \un A} E^{N \un B} F_{\un A \ov A \ov B} F_{\un B}{}^{\ov A \ov B} \partial_{M N}{f} \nn \\ && - 2 E^{\un A}{F_{\un A \ov A \ov B}} E^{\un B}{f} F_{\un B}{}^{\ov A \ov B} +  K_{\un A} {\ov K}_{\ov A} (\frac{\sqrt{2}}{4}  E^{\un A}{E^{M \un B}} F^{\ov A \ov B \ov C} \partial_{M}{F_{\un B \ov B \ov C}} + E^{M \un A} E^{N \un B} F^{\ov A \ov B \ov C} \partial_{M N}{F_{\un B \ov B \ov C}} \nn \\ && + \frac12 E^{\un B}{F_{\un B \ov B \ov C}} E^{\un A}{F^{\ov A \ov B \ov C}} + \frac12 E^{\ov A}{F^{\un A \ov B \ov C}} E^{\un B}{F_{\un B \ov B \ov C}} + \frac{\sqrt{2}}{4}  E^{\un A}{E^{M}{}_{\un B}} F^{\un B \ov B \ov C} \partial_{M}{F_{\ov B \ov C}{}^{\ov A}}  \nn \\ && + \frac{\sqrt{2}}{4}  E^{\un B}{E^{M \ov A}} F_{\un B}{}^{\ov B \ov C} \partial_{M}{F^{\un A}{}_{\ov B \ov C}}  + \frac{\sqrt{2}}{4}  E^{\ov A}{E^{M \un B}} F^{\un A \ov B \ov C} \partial_{M}{F_{\un B \ov B \ov C}}  + E^{M \ov A} E^{N \un B} F_{\un B}{}^{\ov B \ov C} \partial_{M N}{F^{\un A}{}_{\ov B \ov C}} \nn \\ &&
+ E^{M \ov A} E^{N \un B} F^{\un A \ov B \ov C} \partial_{M N}{F_{\un B \ov B \ov C}} + \frac{\sqrt{2}}{4} E^{\un B}{E^{M \un A}} F^{\ov A \ov B \ov C} \partial_{M}{F_{\un B \ov B \ov C}}  + \frac14 E^{\un B}{E^{M \ov A}} F^{\un A \ov B \ov C} \partial_{M}{F_{\un B \ov B \ov C}}  \nn \\ && + \frac{\sqrt{2}}{4} E^{\ov A}{E^{M}_{\un B}} F^{\un B \ov B \ov C} \partial_{M}{F^{\un A}{}_{\ov B \ov C}} - \frac12 E_{\un B}{F^{\un A}{}_{\ov B \ov C}} E^{\un B}{F^{\ov A \ov B \ov C}}  - \frac12 E^{\ov A}{F^{\un B \ov B \ov C}} E^{\un A}{F_{\un B \ov B \ov C}} \nn \\ && + \sqrt{2} E^{\un B}{F_{\un B \ov B}{}^{\ov A}} E^{\un C}{F^{\ov B}{}_{\un C}{}^{\un A}} + E_{\un B}{F_{\un C \ov B}^{\ov A}} E^{\un B}{F^{\ov B \un A \un C}}  + \frac{\sqrt{2}}{4} E^{\un B}{E^{M \un A}} F_{\un B}{}^{\ov B \ov C} \partial_{M}{F_{\ov B \ov C}{}^{\ov A}}  \nn \\ && + E^{M \un B} E^{N \un A} F_{\un B}{}^{\ov B \ov C} \partial_{M N}{F_{\ov B \ov C}{}^{\ov A}} - \frac{1}{\sqrt{2}} E^{\un B}{E^{M \un C}} F_{\un B}{}^{\ov A \ov B} \partial_{M}{F_{\ov B \un C}{}^{\un A}}  - 2 E^{M \un B} E^{N \un C} F_{\un B}{}^{\ov A \ov B} \partial_{M N}{F_{\ov B \un C}{}^{\un A}} \nn \\ && - \frac12 \sqrt{2} E^{\un B}{E^{M}{}_{\un C}} F^{\un C \ov A \ov B} \partial_{M}{F_{\ov B \un B}{}^{\un A}}  + \frac{1}{\sqrt{2}} E^{\un B}{E^{M \un C}} F^{\ov B}{}_{\un B}{}^{\un A} \partial_{M}{F_{\un C \ov B}{}^{\ov A}}  + 2 E^{M \un B} E^{N \un C} F^{\ov B}{}_{\un B}{}^{\un A} \partial_{M N}{F_{\un C \ov B}{}^{\ov A}} \nn \\ && - \frac12 \sqrt{2} E^{\un B}{E^{M}{}_{\un C}} F^{\ov B \un A \un C} \partial_{M}{F_{\un B \ov B}{}^{\ov A}} ) - \frac{1}{\sqrt{2}} E^{\un A}{E^{M \un B}} F_{\un A}{}^{\ov A \ov B} \partial_{M}{F_{\un B \ov A \ov B}}  - 2 E^{M \un A} E^{N \un B} F_{\un A}{}^{\ov A \ov B} \partial_{M N}{F_{\un B \ov A \ov B}} \nn \\ && - \frac{1}{\sqrt{2}}  E^{\un A}{E^{M}{}_{\un B}} F^{\un B \ov A \ov B} \partial_{M}{F_{\un A \ov A \ov B}}  - \frac12 E^{\un A}{F_{\un A \ov A \ov B}} E^{\un B}{F_{\un B}{}^{\ov A \ov B}} + \frac{1}{\sqrt{2}} E^{\un A}{F^{\un B \ov A \ov B}} E_{\un A}{F_{\un B \ov A \ov B}} \nn \\ &&
+ E^{\ov A}{\ov K_{\ov B}} E^{\un A}{K_{\un A}} F^{\un B} F_{\un B \ov A}{}^{\ov B} - \frac12  E^{\un A}{K_{\un B}} E^{\ov A}{\ov K_{\ov A}} F_{\un A}{}^{\ov B \ov C} F^{\un B}{}_{\ov B \ov C} + \frac12  E^{\un A}{\ov K_{\ov A}} E^{\un B}{K_{\un B}} F^{\ov B \ov C \ov A} F_{\un A \ov B \ov C} \nn \\ && 
 +  E^{\ov A}{K_{\un A}} E^{\ov B}{\ov K_{\ov C}} ( - F_{\ov A}{}^{\un B \un A} F_{\un B \ov B}{}^{\ov C} - F_{\ov B}{}^{\un B \un A} F_{\un B \ov A}{}^{\ov C}) +  E^{\un A}{K_{\un B}} E^{\ov A}{\ov K_{\ov B}} ( - F^{\un B} F_{\un A \ov A}{}^{\ov B} + F_{\un A \un C}{}^{\un B} F^{\un C}{}_{\ov A}{}^{\ov B} \nn \\ && 
 - \frac12 F_{\un A \ov A}{}^{\ov C} F^{\un B}{}_{\ov C}{}^{\ov B} + \frac12 F_{\un A}{}^{\ov C \ov B} F^{\un B}{}_{\ov A \ov C}) + \frac12  E^{\un A}{K_{\un B}} E^{\un B}{\ov K_{\ov A}} F^{\ov B \ov C \ov A} F_{\un A \ov B \ov C} \nn \\ && 
 +  E^{\un A}{K_{\un B}} E^{\un C}{\ov K_{\ov A}} (F^{\ov B}{}_{\un C}{}^{\un B} F_{\un A \ov B}{}^{\ov A} + F^{\ov B}{}_{a}{}^{\un B} F_{\un C \ov B}{}^{\ov A}) +  E^{\un A}{\ov K_{\ov A}} E^{\ov B}{K_{\un A}} F^{\un B} F_{\un B \ov B}{}^{\ov A} \nn
\eea}
\footnotesize{
\bea
 && 
 +  E^{\un A}{\ov K_{\ov A}} E^{\ov B}{K_{\un B}} ( - F^{\un B} F_{\un A \ov B}{}^{\ov A} + F_{\un A \un C}{}^{\un B} F^{\un C}{}_{\ov B}{}^{\ov A} - \frac12 F_{\un A \ov B}{}^{\ov C} F^{\un B}{}_{\ov C}{}^{\ov A} + \frac12 F_{\un A}{}^{\ov C \ov A} F^{\un B}{}_{\ov B \ov C}) \nn \\ &&
+  K_{\un A} E^{\ov A}{\ov K_{\ov A}} ( - \frac12 E^{\un B}{F_{\un B \ov B \ov C}} F^{\un A \ov B \ov C} + \frac12 E^{\un B}{F^{\un A}{}_{\ov B \ov C}} F_{\un B}{}^{\ov B \ov C}) +  K_{\un A} E^{\ov A}{\ov K_{\ov B}} ( - E^{\un A}{F_{\un B \ov A}{}^{\ov B}} F^{\un B} + E^{a}{F_{\un B}} F^{\un B}{}_{\ov A}{}^{\ov B} \nn \\ && 
 + \frac12 E^{\un A}{F_{\un B \ov A \ov C}} F^{\un B \ov C \ov B} - \frac12 E^{\un A}{F_{\un B \ov C}{}^{\ov B}} F^{\un B}{}_{\ov A}{}^{\ov C} + E^{\un B}{F_{\un B \ov A}{}^{\ov B}} F^{\un A}  - E^{\ov C}{F_{\un B \ov A}{}^{\ov B}} F_{\ov C}{}^{\un A \un B} \nn \\ && 
 + E^{\un B}{F_{\un C \ov A}{}^{\ov B}} F_{\un B}{}^{\un A \un C} - \frac12 E^{\un B}{F^{\un A}{}_{\ov A \ov C}} F_{\un B}{}^{\ov C \ov B} + E^{\un B}{F^{\un A}} F_{\un B \ov A}{}^{\ov B} + \frac12 E^{\un B}{F^{\un A}{}_{\ov C}{}^{\ov B}} F_{\un B \ov A}{}^{\ov C}) +  K_{\un A} E^{\un A}{\ov K_{\ov A}} (\frac12 F^{\ov B \ov C \ov A} E^{\un B}{F_{\un B \ov B \ov C}} \nn \\ && 
 + \frac12 E^{\un B}{F_{\ov B \ov C}{}^{\ov A}} F_{\un B}{}^{\ov B \ov C}) +  K_{\un A} E^{\un B}{\ov K_{\ov A}} (\frac12 E^{\un A}{F_{\ov B \ov C}{}^{\ov A}} F_{\un B}{}^{\ov B \ov C} - \frac14 E^{\ov A}{F_{\un B \ov B \ov C}} F^{\un A \ov B \ov C} + E^{\un C}{F_{\un C \ov B}{}^{\ov A}} F^{\ov B}{}_{\un B}{}^{\un A} \nn \\ && 
 + \frac34 E^{\ov A}{F^{\un A}{}_{\ov B \ov C}} F_{\un B}{}^{\ov B \ov C} + E^{\un C}{F_{\ov B \un B}{}^{\un A}} F_{\un C}{}^{\ov B \ov A} + E^{\un C}{F_{\ov B \un C}{}^{\un A}} F_{\un B}{}^{\ov B \ov A}) \nn \\ &&
  +  K_{\un A} \partial_{M}{\ov K_{\ov A}} (\frac{\sqrt{2}}{4} F^{\ov B \ov C \ov A} E^{\un A}{E^{M}{}_{\un B}} F^{\un B}{}_{\ov B \ov C}  - \frac{1}{\sqrt{2}} F^{\ov B \ov C \ov A} E^{\un B}{F^{\un A}{}_{\ov B \ov C}} E^{M}{}_{\un B} \nn \\ && 
 + \frac{\sqrt{2}}{4} F^{\ov B \ov C \ov A} E^{\un B}{E^{M \un A}} F_{\un B \ov B \ov C} + \sqrt{2} E^{\un A}{E^{M}_{\ov B}} F^{\un B} F_{\un B}{}^{\ov B \ov A} - \frac{1}{\sqrt{2}} E^{\un B}{E^{M \ov A}} F_{\un B}{}^{\ov B \ov C} F^{\un A}{}_{\ov B \ov C}  \nn \\ && 
 - \sqrt{2} E^{\un B}{E^{M}{}_{\ov B}} F^{\un A} F_{\un B}{}^{\ov B \ov A} - \sqrt{2} E^{\ov B}{E^{M}{}_{\ov C}} F_{\ov B}^{\un B \un A} F_{\un B}{}^{\ov C \ov A}  + \frac{1}{\sqrt{2}} E^{\un B}{E^{M}_{\ov B}} F_{\un B}{}^{\ov C \ov B} F^{\un A}{}_{\ov C}{}^{\ov A}  \nn \\ && 
 - \frac{1}{\sqrt{2}} E^{\un B}{E^{M}{}_{\ov B}} F_{\un B}{}^{\ov C \ov A} F^{\un A}{}_{\ov C}{}^{\ov B} - \frac{1}{\sqrt{2}} E^{\un B}{E^{M}{}_{\un C}} F^{\ov B \un A \un C} F_{\un B \ov B}{}^{\ov A} + \frac{1}{\sqrt{2}} E^{\un B}{E^{M}{}_{\un C}} F^{\ov B}{}_{\un B}{}^{\un A} F^{\un C}{}_{\ov B}{}^{\ov A}  \nn \\ && 
 + \sqrt{2} E^{\un B}{F_{\un C \ov B}{}^{\ov A}} E^{M}{}_{\un B} F^{\ov B \un A \un C} + \sqrt{2} E^{\un B}{E^{M}{}_{\ov B}} F_{\un B \un C}{}^{\un A} F^{\un C \ov B \ov A} ) +  K_{\un A} \partial_{M N}{\ov K_{\ov A}} (F^{\ov B \ov C \ov A} E^{M \un B} E^{N \un A} F_{\un B \ov B \ov C} \nn \\ &&
+ 2 E^{M \ov B} E^{N \un A} F^{\un B} F_{\un B \ov B}{}^{\ov A} - E^{M \ov A} E^{N \un B} F_{\un B}{}^{\ov B \ov C} F^{\un A}{}_{\ov B \ov C} - 2 E^{M \ov B} E^{N \un B} F^{\un A} F_{\un B \ov B}{}^{\ov A} \nn \\ && 
 - 2 E^{M \ov B} E^{N \ov C} F_{\ov B}{}^{\un B \un A} F_{\un B \ov C}{}^{\ov A} - E^{M \ov B} E^{N \un B} F_{\un B \ov B}{}^{\ov C} F^{\un A}{}_{\ov C}{}^{\ov A} + E^{M \ov B} E^{N \un B} F_{\un B}{}^{\ov C \ov A} F^{\un A}{}_{\ov B \ov C} \nn \\ && 
 + 2 E^{M \un B} E^{N \un C} F^{\ov B}{}_{\un B}{}^{\un A} F_{\un C \ov B}{}^{\ov A} + 2 E^{M \ov B} E^{N \un B} F_{\un B \un C}{}^{\un A} F^{\un C}{}_{\ov B}{}^{\ov A}) +  \ov K_{\ov A} E^{\ov B}{K_{\un A}} (E^{\un A}{F_{\un B}} F^{\un B}{}_{\ov B}{}^{\ov A} \nn \\ && 
 + E^{\un B}{F^{\un A}} F_{\un B \ov B}{}^{\ov A} + \frac12 E^{\un B}{F^{\un A}{}_{\ov B \ov C}} F_{\un B}{}^{\ov A \ov C} - \frac12 E^{\un A}{F_{\un B \ov B \ov C}} F^{\un B \ov A \ov C} - E^{\un A}{F_{\un B \ov B}{}^{\ov A}} F^{\un B} - E^{\un B}{F_{\un C \ov B}{}^{\ov A}} F_{\un B}{}^{\un C \un A} \nn \\ && 
 + E^{\un B}{F_{\un B \ov B}{}^{\ov A}} F^{\un A} - \frac12 E^{\un A}{F_{\un B \ov C}{}^{\ov A}} F^{\un B}{}_{\ov B}{}^{\ov C} + \frac12 E^{\un B}{F^{\un A}{}_{\ov C}{}^{\ov A}} F_{\un B \ov B}{}^{\ov C} + E^{\ov C}{F_{\un B \ov B}{}^{\ov A}} F_{\ov C}{}^{\un B \un A}) \nn \\ && 
 +  \ov K_{\ov A} E^{\un A}{K_{\un A}} (\frac12 F^{\ov A \ov B \ov C} E^{\un B}{F_{\un B \ov B \ov C}} + \frac12 E^{\un B}{F_{\ov B \ov C}{}^{\ov A}} F_{\un B}{}^{\ov B \ov C}) +  \ov K_{\ov A} E^{\un A}{K_{\un B}} ( - \frac14 E^{\ov A}{F_{\un A \ov B \ov C}} F^{\un B \ov B \ov C} \nn \\ && + \frac34 E^{\ov A}{F^{\un B}{}_{\ov B \ov C}} F_{\un A}{}^{\ov B \ov C} 
 - E^{\un C}{F_{\ov B \un A}{}^{\un B}} F_{\un C}{}^{\ov A \ov B} - E^{\un C}{F_{\ov B \un C}{}^{\un B}} F_{\un A}{}^{\ov A \ov B} + \frac12 E^{\un B}{F_{\ov B \ov C}{}^{\ov A}} F_{\un A}{}^{\ov B \ov C} + E^{\un C}{F_{\un C \ov B}{}^{\ov A}} F^{\ov B}{}_{\un A}{}^{\un B}) \nn \\ && 
 +  \ov K_{\ov A} \partial_{M}{K_{\un A}} ( - \frac{1}{\sqrt{2}} F^{\ov A \ov B \ov C} E^{\un B}{F^{\un A}{}_{\ov B \ov C}} E^{M}{}_{\un B}   + \frac{\sqrt{2}}{4} F^{\ov B \ov C \ov A} E^{\un A}{E^{M}{}_{\un B}} F^{\un B}{}_{\ov B \ov C} + \frac{\sqrt{2}}{4} F^{\ov B \ov C \ov A} E^{\un B}{E^{M \un A}} F_{\un B \ov B \ov C}  \nn \\ && 
 - \sqrt{2} E^{\un A}{E^{M}{}_{\ov B}} F^{\un B} F_{\un B}{}^{\ov A \ov B}  + \sqrt{2} E^{\un B}{E^{M}{}_{\ov B}} F^{\un A} F_{\un B}{}^{\ov A \ov B} + \frac{1}{\sqrt{2}} E^{\un B}{E^{M}{}_{\un C}} F^{\ov B \un C \un A} F_{\un B \ov B}{}^{\ov A} + \frac{1}{\sqrt{2}} E^{\un B}{E^{M}{}_{\un C}} F^{\ov B}{}_{\un B}{}^{\un A} F^{\un C}{}_{\ov B}{}^{\ov A} \nn \\ && 
 - \sqrt{2} E^{\un B}{E^{M}{}_{\ov B}} F_{\un B \un C}{}^{\un A} F^{\un C \ov A \ov B} + \sqrt{2} E^{\ov B}{E^{M}{}_{\ov C}} F_{\ov B}{}^{\un B \un A} F_{\un B}{}^{\ov A \ov C} + \frac{1}{\sqrt{2}} E^{\un B}{E^{M}{}_{\ov B}} F_{\un B}{}^{\ov C \ov B} F^{\un A}{}_{\ov C}{}^{\ov A} \nn \\ && 
 - \frac{1}{\sqrt{2}} E^{\un B}{E^{M}{}_{\ov B}} F_{\un B}{}^{\ov C \ov A} F^{\un A}{}_{\ov C}{}^{\ov B} - \sqrt{2} E^{\un B}{F_{\un C \ov B}{}^{\ov A}} E^{M}{}_{\un B} F^{\ov B \un C \un A}) +  \ov K_{\ov A} \partial_{M N}{K_{\un A}} (F^{\ov B \ov C \ov A} E^{M \un B} E^{N \un A} F_{\un B \ov B \ov C} \nn \\ && 
 + 2 E^{M \ov B} E^{N \un A} F^{\un B} F_{\un B \ov B}{}^{\ov A} - 2 E^{M \ov B} E^{N \un B} F^{\un A} F_{\un B \ov B}{}^{\ov A} + 2 E^{M \un B} E^{N \un C} F^{\ov B}{}_{\un B}{}^{\un A} F_{\un C \ov B}{}^{\ov A} \nn \\ && 
 + 2 E^{M \ov B} E^{N \un B} F_{\un B \un C}{}^{\un A} F^{\un C}{}_{\ov B}{}^{\ov A} - 2 E^{M \ov B} E^{N \ov C} F_{\ov B}{}^{\un B \un A} F_{\un B \ov C}{}^{\ov A} - E^{M \ov B} E^{N \un B} F_{\un B \ov B}{}^{\ov C} F^{\un A}{}_{\ov C}{}^{\ov A} \nn \\ && + E^{M \ov B} E^{N \un B} F_{\un B}{}^{\ov C \ov A} F^{\un A}{}_{\ov B \ov C}) \, ,
\eea}

\footnotesize{
\bea
T_{3} & = &  E^{\ov A}{K_{\un A}} ( - \frac{1}{\sqrt{2}} E^{\un B}{E^{M \un A}} F_{\un B \ov A}{}^{\ov B} \partial_{M}{\ov K_{\ov B}}  - 2 E^{M \un A} E^{N \un B} \ov K^{\ov B} \partial_{M N}{F_{\un B \ov A \ov B}} - 2 E^{M \un A} E^{N \un B} F_{\un B \ov A}{}^{\ov B} \partial_{M N}{\ov K_{\ov B}} \nn \\ &&
- \frac{1}{\sqrt{2}} E^{\un A}{E^{M}{}_{\un B}} F^{\un B}{}_{\ov A}{}^{\ov B} \partial_{M}{\ov K_{\ov B}}  + \sqrt{2} E^{\un B}{F^{\un A}{}_{\ov A}{}^{\ov B}} E_{\un B}{\ov K_{\ov B}}) \nn \\ && 
+  E^{\ov A}{\ov K_{\ov B}} ( - \frac{1}{\sqrt{2}} E^{\un A}{E^{M \un B}} F_{\un A \ov A}{}^{\ov B} \partial_{M}{K_{\un B}}  - 2 E^{M \un A} E^{N \un B} F_{\un A \ov A}{}^{\ov B} \partial_{M N}{K_{\un B}} - \frac{1}{\sqrt{2}} E^{\un A}{E^{M}{}_{\un B}} F^{\un B}{}_{\ov A}{}^{\ov B} \partial_{M}{K_{\un A}}  \nn \\ && 
+ \sqrt{2} E^{\un A}{F^{\un B}{}_{\ov A}{}^{\ov B}} E_{\un A}{K_{\un B}}) +  E^{\un A}{K_{\un A}} ( - 2 E^{M \ov A} E^{N \un B} F_{\un B \ov A}{}^{\ov B} \partial_{M N}{\ov K_{\ov B}} - \sqrt{2} E^{\un B}{E^{M}{}_{\ov A}} F_{\un B}{}^{\ov A \ov B} \partial_{M}{\ov K_{\ov B}}) \nn \\ && 
+  E^{\un A}{K_{\un B}} ( - 2 E^{M \ov A} E^{N \un B} F_{\un A \ov A}{}^{\ov B} \partial_{M N}{\ov K_{\ov B}} - \sqrt{2} E^{\un B}{E^{M}{}_{\ov A}} F_{\un A}{}^{\ov A \ov B} \partial_{M}{\ov K_{\ov B}}) \nn \\ && 
+  E^{\un A}{\ov K_{\ov A}} ( - 2 E^{M \ov B} E^{N \un B} F_{\un A \ov B}{}^{\ov A} \partial_{M N}{K_{\un B}} - 2 E^{M \ov B} E^{N \un B} F_{\un B \ov B}{}^{\ov A} \partial_{M N}{K_{\un A}} - \sqrt{2} E^{\un B}{E^{M}{}_{\ov B}} F_{\un B}{}^{\ov B \ov A} \partial_{M}{K_{\un A}} \nn \\ && 
- \sqrt{2} E^{\un B}{E^{M}{}_{\ov B}} F_{\un A}{}^{\ov B \ov A} \partial_{M}{K_{\un B}}) +  K_{\un A} ( - 2 E^{\ov A}{\ov K^{\ov B}} E^{M \un A} E^{N \un B} \partial_{M N}{F_{\un B \ov A \ov B}} \nn \\ && 
- 2 E^{\un A}{E^{M}{}_{\ov A}} E^{N \un B} F_{\un B}{}^{\ov A \ov B} \partial_{M N}{\ov K_{\ov B}} - E^{\un A}{E_{M \un B}} F^{\un B \ov A \ov B} \partial_{N}{\ov K_{\ov B}} \partial^{M}{E^{N}{}_{\ov A}} \nn \\ &&
- E^{\un A}{E^{M}{}_{\un B}} E^{N \ov A} F^{\un B}{}_{\ov A}{}^{\ov B} \partial_{M N}{\ov K_{\ov B}} - \sqrt{2} E^{\un A}{E^{M \ov A}} E^{\un B}{F_{\un B \ov A}{}^{\ov B}} \partial_{M}{\ov K_{\ov B}}  + 2 \sqrt{2} E^{M \un B} E^{N \un A} F_{\un B}{}^{\ov A \ov B} \partial^{P}{\ov K_{\ov A}} \partial_{M N}{E_{P \ov B}} \nn \\ && 
- 2 E^{\un B}{E^{M}{}_{\ov A}} E^{N \un A} F_{\un B}{}^{\ov A \ov B} \partial_{M N}{\ov K_{\ov B}} - 2 \sqrt{2} E^{M \ov A} E^{N \un B} E^{P \un A} F_{\un B \ov A}{}^{\ov B} \partial_{M N P}{\ov K_{\ov B}}  - 2 E^{\un B}{F_{\un B \ov A}{}^{\ov B}} E^{M \ov A} E^{N \un A} \partial_{M N}{\ov K_{\ov B}} \nn \\ &&
- F_{\un B}{}^{\ov A \ov B} E^{\un B}{E_{M}{}^{\un A}} \partial_{N}{\ov K_{\ov B}} \partial^{M}{E^{N}{}_{\ov A}} - E^{\un B}{E^{M \un A}} E^{N \ov A} F_{\un B \ov A}{}^{\ov B} \partial_{M N}{\ov K_{\ov B}} + 2 E^{\un B}{E^{M \ov A}} E_{\un B}{F^{\un A}{}_{\ov A}{}^{\ov B}} \partial_{M}{\ov K_{\ov B}} \nn \\ && 
+ 2 E^{M \ov A} E^{N}{}_{\un B} E^{\un B}{F^{\un A}{}_{\ov A}{}^{\ov B}} \partial_{M N}{\ov K_{\ov B}})  +  \ov K_{\ov A} (E^{\un A}{E_{M}{}^{\un B}} F_{\un A}{}^{\ov A \ov B} \partial_{N}{K_{\un B}} \partial^{M}{E^{N}{}_{\ov B}} \nn \\ &&
+ 2 \sqrt{2} E^{M \un A} E^{N \un B} F_{\un A}{}^{\ov A \ov B} \partial^{P}{K_{\un B}} \partial_{M N}{E_{P \ov B}} + E^{\un A}{E_{M \un B}} F^{\un B \ov A \ov B} \partial_{N}{K_{\un A}} \partial^{M}{E^{N}{}_{\ov B}} \nn \\ &&
- E^{\un A}{E^{M \un B}} E^{N \ov B} F_{\un A \ov B}{}^{\ov A} \partial_{M N}{K_{\un B}} + 2 E^{\un A}{E^{M}{}_{\ov B}} E^{N \un B} F_{\un A}{}^{\ov A \ov B} \partial_{M N}{K_{\un B}} + 2 E^{\un A}{E^{M}{}_{\ov B}} E^{N \un B} F_{\un B}{}^{\ov A \ov B} \partial_{M N}{K_{\un A}} \nn \\ && 
- 2 \sqrt{2} E^{M \ov B} E^{N \un A} E^{P \un B} F_{\un A \ov B}{}^{\ov A} \partial_{M N P}{K_{\un B}}  - E^{\un A}{E^{M}{}_{\un B}} E^{N \ov B} F^{\un B}{}_{\ov B}{}^{\ov A} \partial_{M N}{K_{\un A}} - \sqrt{2} E^{\un A}{E^{M \ov B}} E^{\un B}{F_{\un B \ov B}{}^{\ov A}} \partial_{M}{K_{\un A}}  \nn \\ && 
- 2 E^{\un A}{F_{\un A \ov B}{}^{\ov A}} E^{M \ov B} E^{N \un B} \partial_{M N}{K_{\un B}} - 2 E^{\un A}{E^{M \ov B}} E_{\un A}{F^{\un B \ov A}{}_{\ov B}} \partial_{M}{K_{\un B}} + 2 E^{M \ov B} E^{N}{}_{\un A} E^{\un A}{F^{\un B}{}_{\ov B}{}^{\ov A}} \partial_{M N}{K_{\un B}}) \nn \\ && 
-  E^{\ov A}{\ov K_{\ov B}} E^{\un A}{K_{\un A}} E^{\un B}{F_{\un B \ov A}{}^{\ov B}} -  E^{\un A}{\ov K_{\ov A}} E^{\ov B}{K_{\un A}} E^{\un B}{F_{\un B \ov B}{}^{\ov A}} +  K_{\un A} E^{\ov A}{\ov K_{\ov B}} ( - \frac{1}{\sqrt{2}} E^{\un A}{E^{M \un B}} \partial_{M}{F_{\un B \ov A}{}^{\ov B}}  \nn \\ && 
- \frac{1}{\sqrt{2}} E^{\un B}{E^{M \un A}} \partial_{M}{F_{\un B \ov A}{}^{\ov B}} ) +  \ov K_{\ov A} E^{\ov B}{K_{\un A}} ( - \frac{1}{\sqrt{2}} E^{\un A}{E^{M \un B}} \partial_{M}{F_{\un B \ov B}{}^{\ov A}}  - \frac{1}{\sqrt{2}} E^{\un B}{E^{M \un A}} \partial_{M}{F_{\un B \ov B}{}^{\ov A}}) \, .
\eea}

\end{document}